\documentclass[journal]{IEEEtran}

\ifCLASSINFOpdf
\else
\fi

\usepackage{etex}
\usepackage{algorithmicx}
\usepackage{algpseudocode}
\usepackage[ruled,vlined]{algorithm2e}
\SetKwRepeat{Do}{do}{while}

\usepackage{amsmath,amssymb,amsthm}
\usepackage{balance}
\usepackage{mathtools}
\usepackage[justification=centering]{caption}
\usepackage[noadjust]{cite}
\usepackage{array}
\usepackage{graphicx}
\usepackage{epstopdf}
\usepackage{setspace}
\usepackage[export]{adjustbox}
\usepackage{ctable}
\usepackage{enumerate}
\usepackage{ragged2e}
\usepackage{amssymb}
\usepackage{multirow}
\usepackage{tabularx,tabulary}
\usepackage{nicefrac}		
\usepackage[caption=false,font=footnotesize,labelformat=simple]{subfig} 
\usepackage[flushleft]{threeparttable}
\usepackage[bookmarks=false]{hyperref}
\usepackage{rotating}
\usepackage{tabularx,ragged2e,booktabs}
\newcolumntype{C}[1]{>{\Centering}m{#1}}

\usepackage{amssymb}
\usepackage{pifont}
\usepackage{mathtools}
\newcommand{\cmark}{\ding{51}}%
\newcommand{\xmark}{\ding{55}}%

\usepackage{authblk}
\usepackage{verbatim}

\DeclarePairedDelimiter{\floor}{\lfloor}{\rfloor}

\usepackage{xcolor}
\usepackage{tcolorbox}
\newtcbox{\highlight}[0]{boxsep=0pt,left=0pt,top=0pt,bottom=0pt,right=0pt,boxrule=0pt,arc=0pt,auto outer arc,colback=green,width=15cm}

\definecolor{dark-blue}{RGB}{0,70,127}
\newcommand\bt[1]{\textcolor{dark-blue}{#1}}

\usepackage[final]{pdfpages}
\usepackage{pifont}
\usepackage{framed}
\usepackage{soul}
\usepackage{booktabs}
\usepackage{multirow}
\usepackage{mathtools}

\begin{document}


\title{Fuzzing+Hardware Performance Counters-Based Detection of Algorithm Subversion Attacks on Post-Quantum Signature Schemes}
\author{Animesh Basak Chowdhury, Anushree Mahapatra,
        Deepraj Soni, and
        Ramesh Karri,~\IEEEmembership{Fellow,~IEEE}
\thanks{A.B. Chowdhury, A. Mahapatra, D. Soni, and R. Karri are with the Dept.
of Electrical and Computer Engineering, New York University, Brooklyn,
NY, 11201 USA. E-mail: \{am11019, abc586, dss545, rkarri\}@nyu.edu
}

}

\maketitle

\begin{abstract}
NIST is standardizing Post Quantum Cryptography (PQC) algorithms that are resilient to the computational capability of quantum computers. Past works show malicious subversion with cryptographic software (\textit{algorithm subversion attacks}) that weaken the implementations. We show that  PQC digital signature codes can be subverted in line with  previously reported flawed implementations~\cite{debssl2008,dual_ec_2016} that generate verifiable, but less-secure signatures, demonstrating the risk of such attacks. Since, all processors have built-in Hardware Performance Counters (HPCs), there exists a body of work proposing a low-cost Machine Learning (ML)-based integrity checking of software using HPC fingerprints. However, such HPC-based approaches may not detect subversion of PQC codes. A miniscule percentage of qualitative inputs when applied to the PQC codes improve this accuracy to 98\%. We propose grey-box fuzzing as a pre-processing step to obtain inputs to aid the HPC-based method.
\end{abstract}


\IEEEpeerreviewmaketitle

\section{Introduction}
 
\textit{Algorithm Subversion Attacks} (\emph{ASA}) on cryptographic software deployed for public use is an important class of attacks on cryptosystems besides the well-known side-channel and fault attacks~\cite{klepto-crypto96,asa-crypto2014,debssl2008}. ASA is also known as \textit{Kleptography}~\cite{klepto-crypto96}. ASAs weaken cryptography implementation without user knowledge. The subverted software leaks all (or part) of the secret key during message encryption and signature generation. The attacker can recover the secret key of any party that uses such a subverted system. Flawed implementations by software developers introduce vulnerabilities, which when discovered can be exploited~\cite{jafargholi2015tamper,boneh2001importance}.

The main challenge in detecting ASAs is that the outputs generated by the subverted crypto software are computationally indistinguishable from those generated by trusted implementations. The subversion is  due to the low resilience to crypt-analysis, fragility of implementation due to bad randomness, reuse of nonces, and leakage of sensitive data via side-channels. \emph{ASA} was first studied and systematically explored by Young \textit{et al.}~\cite{klepto-crypto96}, where they showed feasibility of such attacks on RSA (they called it the SETUP attack). Bellare \textit{et al.} extended this attack to symmetric encryption standards~\cite{asa-crypto2014}. Recent works show that Post-Quantum Cryptography implementations are vulnerable to such attacks~\cite{lwe-asa-2020,lattice-asa-2020,stegnano2017}.

This paper will study detection of \emph{subversions} in PQC implementations. State-of-the-art literature suggests two pathways to evade \emph{ASA}: 1) design of subversion-resilient algorithms and 2) detection. Subversion-resilient design of cryptography algorithms were proposed in \cite{clipto2019,subvert-resilient}. Designing subversion-resilient algorithms requires cryptanalysis of the algorithm. Complementing this approach, we propose a scheme to check if the subverted software implementations leave unique micro-architectural traces during execution different from the traces of a trusted PQC software implementation. We will investigate Machine Learning (ML)-based detectors using Hardware Performance Counters (HPC).

HPCs are low-cost performance monitors built into all processors and can be reused at no extra cost. HPCs have been used as light-weight instrumentation tools in embedded systems to monitor the run-time behaviour of software~\cite{malone2011hardware,alam2017performance,wang2016sigdrop,krishnamurthy2019anomaly}. HPC-based integrity check has been shown to be a successful and effective low-cost detection solution~\cite{malone2011hardware,wang2016sigdrop}. We started with using ML-based detection technique to classify a subverted PQC implementation using run-time HPC traces. Unfortunately, our experimental findings in section~\ref{sec:findingsSOTA} show that 95\% of HPC traces for a subverted implementation are similar to those for a trusted implementation. Clearly, vanilla ML-based detectors cannot classify the run-time HPC traces of subverted implementation. We overcame this limitation by using Greybox fuzzing to determine the unique input to the PQC implementations that create distinguishable HPC traces.
Contributions of this study are three-fold:
\begin{enumerate}
\item We show that PQC implementations can be  subverted~\cite{debssl2008,dual_ec_2016,heartbleed,freak,poodle} in at least three ways: (i) tampering with PQC security parameters, (ii) tweaking the Random Number generators, and (iii) subverting the hash functions. We show that ML-based classifiers that monitor HPC traces on random inputs cannot detect the subversions.
\item We adopt Greybox fuzzing to discover inputs that yield discernible HPC signatures of the PQC codes to distinguish subverted PQC implementations from the original.
\item We build an ML classifier that performs temporal and spatial integrity checks using discernible HPC signatures on inputs derived using Greybox fuzzing of PQC implementations.
\end{enumerate}

The paper is organized as follows: Section~\ref{sec:prelims} lays out the threat model and preliminaries on HPCs and PQC signature algorithms. Section~\ref{sec:anomalies} describes three types of subversions introduced into PQC signature algorithms. Section~\ref{sec:limitationsOfSOTA} outline the challenges of using vanilla ML-based HPC detection scheme. Section~\ref{sec:methodology} shows greybox fuzzing guided input generation to aid ML detection methodology using HPC signatures. Section~\ref{sec:expresults} discusses the  experimental results. Section~\ref{sec:relatedwork} reviews the related work and finally, Section~\ref{sec:conclusion} presents the conclusions of this study.
\section{Preliminaries}
\label{sec:prelims}

\subsection{Threat Model}
\label{sec:threat}
Our threat model is in line with the setting proposed in \cite{threatModel}. We have four players in the setting: \textit{Saboteur}, \textit{Victim}, \textit{Attacker}, and \textit{Defender}(\autoref{tab:threatModel}). The goal of the \textit{Saboteur} is to stealthily weaken the PQC software, which may later be exploited by an \textit{Attacker} to target specific user(s) and recover their  secret keys or sensitive information. The \textit{victim} is the user who deploys the weakened PQC software. The \textit{defender} is the algorithm-level designer of the PQC software. The goal of the \textit{Defender} is to ensure that \textit{Victim} can easily check the integrity of the PQC software implementations. We encourage the reader to go through \cite{threatModel} for more details. We motivate our threat model using the classic FREAK~\cite{freak} attack, where the \textit{Saboteur} was allegedly the National Security Agency (NSA). They intentionally weakened the OpenSSL protocols by restricting to use RSA\_export key size below 512 bits. This can be easily decrypted by number field sieve factorization algorithm  using modest computing resources. NIST in its role of a \textit{Defender} has to ensure that the \textit{victim} can check for weaknesses in implementations. We assume PQC implementations in TLS/SSL libraries can be subverted intentionally or accidentally by software developers.

\begin{table}[t!]
\centering
\caption{Players in Threat model\cite{threatModel}}
\begin{tabular}{ll}
\hline \hline
\textbf{Player} & \textbf{Goal}     \\ \hline \hline
Saboteur & \begin{tabular}[c]{@{}l@{}}Adds weakness to crypto \\ implementation (Software developer)\end{tabular} \\
Attacker & \begin{tabular}[c]{@{}l@{}}Extract secret information \\from weakened signature (NSA)\end{tabular}\\
Victim & \begin{tabular}[c]{@{}l@{}}Uses subverted crypto\\ implementation (User)\end{tabular}\\
Defender & \begin{tabular}[c]{@{}l@{}}Protects integrity of crypto \\ implementation (NIST)\end{tabular}\\
\hline \hline
\end{tabular}
\label{tab:threatModel}
\end{table}

\subsection{Motivation}
The core research question we ask in our work is: \textit{How can a user trust the implementation of crypto APIs from third-party developers (possible Saboteur)?}
Our threat model shows that the \textit{Attacker} with the help of third-party crypto software developers (\textit{Saboteur}) can perform backdoor injection/subvert cryptography algorithm leaking secret key information through output.
There are multitude reasons leading to failure of existing techniques to solve the problem. 
\begin{itemize}
    \item Third-party developers generally provide static checksums for users to trust libraries are from authentic sources. However, this defense is moot since the source corrupts the implementation.
    \item Honest implementations are available from Defender but are not user-friendly APIs (support for debugging, parallelism). One can use such implementation and tune it according to the needs. This directly mitigate any threats from third-party developers. However, such scenario is out of scope from our work.
    \item We show seeds provided by Defender are random and do not fully explore the state-space of crypto implementations. Thus, validation based on random test-inputs is incomplete and backdoor may still be hidden.
\end{itemize}

We propose the defender to release dynamic/runtime signatures generated by running fuzz-guided inputs on PQC implementations. For versatility across different architectures (arm, x86), we assume the defender creates the binary for all notable architectures with predefined libraries and compiler versions. The defender/victim can train a ML-based model using runtime HPC signatures generated by fuzzed input for detecting anomalous behaviour in PQC implementations. We believe our work will help NIST's ongoing Automated Cryptographic Validation Testing(ACVT) program\cite{skeller_2018}.

\subsection{Overview of PQC Digital Signature Algorithms}
\begin{table}[t!]
\centering
\caption{NIST PQC Digital Signature candidates\cite{nistpqc}}
\begin{tabular}{ll}
\hline \hline
\textbf{Type}      & \textbf{PQC Digital Signature}     \\ \hline \hline
Lattice-based      & Dilithium, Falcon \\
Symmetric-based    & SPHINCS+ (alternate candidate)
\\
Multivariate-based & 
Rainbow\\ \hline \hline
\end{tabular}

\label{fig:pqctaxonomy}
\end{table}
NIST is standardizing quantum-resistant a.k.a post quantum public key cryptography in two main classes: 1. post quantum digital signature (DS) schemes and 2. post quantum Key Encapsulation Mechanisms (KEM). In current work, we study impact of algorithm subversion attacks on PQC DS schemes although these can be similarly extended to KEM algorithms. The DS schemes authenticate the identity of the signatory~\cite{nistpqc} and detect unauthorized modifications to data. There are three main modules of DS: 1) \textit{Keypair} generation: The signatory generates a pair of keys: secret key ($s_k$) and public key ($p_k$) with seed as input. 2) \textit{Signature} generation: The signatory signs a message($m$) with the secret key (\textit{sign($s_k$,$m$)}). 3) \textit{Signature} verification: The authenticator receives the signed message and verifies the signatory using \textit{signVerify($p_k$,$S_{m}$)}. We target the DS algorithms from NIST PQC round 3 submission as summarized in Table \ref{fig:pqctaxonomy}.
The DS candidates are classified into 3 classes based on the underlying mechanisms. Our work demonstrates on all three types. 
\begin{itemize}
    \item \textbf{Lattice-based DS} builds on the hardness of shortest vector problem (SVP) and takes polynomial time in quantum computers.
    \item \textbf{Multi-variate DS} solves Multi-variate Polynomial(MVP) algorithm in finite field and has NP-hard complexity. They are light-weight due to their short signatures.
    \item \textbf{Symmetric-based DS} resists quantum computer attacks based on the security of the underlying cryptographic hash functions.
\end{itemize}

\subsection{Hardware Performance Counters (HPC)}
HPCs are special-purpose registers built into the performance monitoring unit of all processors. HPCs store counts of software and hardware events in the processor. Every HPC provides information about the micro-architectural state and events in different parts of the processor. Example HPCs include events like cache misses and branch mis-predictions. HPC-based monitoring incurs minimal time overhead with zero hardware cost. Further, HPC-based monitoring does not require any extra  modifications to the monitored code.
The HPCs differ from one processor to the next~\cite{tang2014unsupervised}. For example, HPCs in ARM Cortex A8 processor count Level-2 Data Cache misses, while HPCs in AMD x86 processor count the cumulative misses in Level-2 caches.  Table~\ref{tab:HPCeventsARM} tabulates the list of HPCs from ARM Cortex A8 processor that we use. 

\begin{table}[!htbp]
\caption{HPCs and hardware-level events}
\begin{center}
\begin{tabular}{ll}
\hline \hline
\multicolumn{1}{c}{\textbf{HPC}} & \textbf{Hardware events} \\ \hline \hline
CYCLES   & CPU Cycles    \\
L2-TCM  & L2 Total Cache Misses \\
BR-MSP  & Branch  Mispredictions \\ 
L1-ICM   & L1 Instruction Cache Misses   \\
L1-DCA  & L1 Data Cache Accesses  \\
L2-DCA  & L2 Data Cache Accesses    \\
L1-DCM & L1 Data Cache Misses \\
L2-DCM  & L2 Data Cache Misses \\ \hline \hline
\end{tabular}
\label{tab:HPCeventsARM}
\end{center}
\end{table}

\subsection{HPC Signatures}
\begin{figure}[htbp!]
    \centering
    \includegraphics[width=0.8\columnwidth]{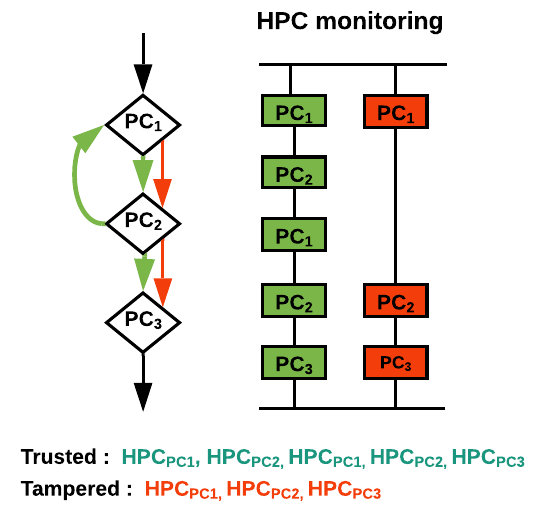}
    \caption{Consider \textit{crypto\_sign()} of Dilithium with program-checkpoints. $PC_1$, $PC_2$ are hit different number of times for trusted (green) and tampered codes (red) yielding different $HPC$ signatures.}
    \label{fig:programcfg}
\end{figure}

HPC signature of a PQC code denotes the HPC values obtained during execution. We use HPC signatures to fingerprint run-time behavior of a PQC code in two ways: The \textbf{time-series based signature} captures the temporal variations of the HPC values during the PQC code execution for a seed input and message. \textbf{Program-checkpoint (PC) based signature}: A PC is a location in the code where HPCs are monitored. Fig.~\ref{fig:programcfg} shows HPC collection across multiple PCs. PC-based HPC signature captures spatial variations of HPC values at different PCs by executing PQC code with seed inputs and message. Tampering with the PQC code results in deviation from the trusted HPC signature.

HPC signatures represent dynamic hashes of PQC DS implementations. Ideally, HPC signatures of a PQC implementation at any given time are cumulative of counts of the monitored HPCs. In reality, when monitoring HPCs at runtime, system noise becomes a factor. Thus, we form HPC signatures  by deriving statistical measures from multiple readings of the HPCs monitored for a period of time. The HPC signatures are then reproducible and deterministic. At any time, when an attacker maliciously modifies the code, the HPC signatures will vary from the pre-computed signatures. Section 7 has a detailed discussion of HPC signatures for PQC DS algorithms.

HPC-based techniques complement the static integrity verification techniques that compute the hashes of the program executable at program installation time and periodically during the program execution~\cite{fiskiran2004runtime}. 

\section{ASA on PQC Digital Signature Codes}
\label{sec:anomalies}
\begin{figure}[!htb]
    \centering
   \includegraphics[width=\columnwidth]{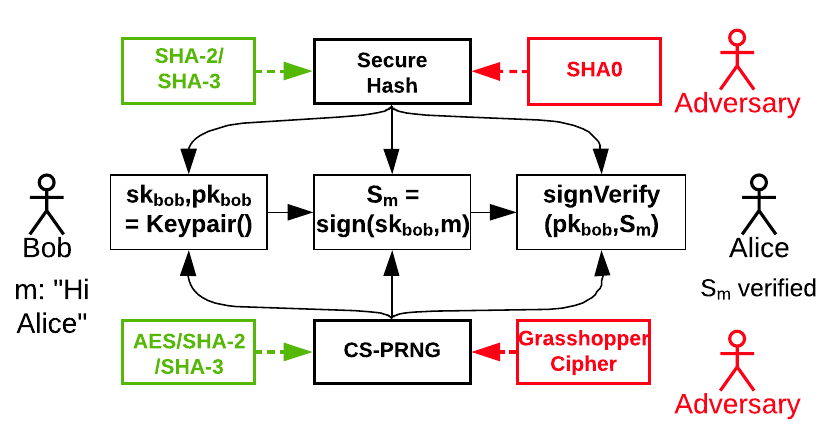}
    \caption{Pseudo-RNG: ASA on Hashes used in PQC DS implementations.
    }
    \label{fig:pqcanomaly}
\end{figure}

In this section, we describe three types of algorithm subversion attacks (ASA) on PQC DS codes to reduce their security strength: 
 \textit{Subverting the random number generator} (PRNG), \textit{Subverting the Hash function} (HASH), \textit{Subverting the Security Parameters} (SPARAM). Our choices of attack for \textit{PRNG} and \textit{HASH} were motivated by software-based attacks FREAK\cite{freak}. PRNG and HASH modifications impact all the cryptographic modules of signature algorithms (\autoref{fig:pqcanomaly}) and generated signature were validated at verification end (Although, an honest verifier showed the signature is corrupted).
 
S-PARAM attacks introduce  vulnerabilities by tampering security parameters. These vulnerabilities are documented by the signature algorithm designers they can be exploited by attacker. We aim to detect subversions in PQC signatures (and not designing them in first place).

 \subsection{ASA on Pseudo Random Number Generators (PRNG)}
A good pseudo-random number generator (PRNG) is essential for all PQCs. In NIST submissions, the PQC DS candidates use block ciphers like AES-256 and cryptographic hashes like SHA-3 to implement PRNGs. These PRNGs underlie all three modules in a PQC DS code namely, key-pair generation, signature creation and verification.
As shown in Fig.~\ref{fig:pqcanomaly}, one can subvert the PRNGs by replacing secure hashes and block ciphers with less secure variants. The adversary reduces the entropy of the output of the PRNG and make it less secure in line with Debian openssl threat\cite{debssl2008}. Table~\ref{tab:rngUsageinPQCDS} shows implementations of PRNG in the PQC digital signatures.

\begin{table}[h]
\resizebox{\columnwidth}{!}{
\begin{tabular}{lcccc}
\hline \hline
\multirow{2}{*}{\textbf{ PRNG}} & \multicolumn{4}{c}{\textbf{PQC DS Algorithms}}    \\ \cline{2-5} 
& Dilithium & Falcon & Rainbow & SPHINCS+ \\ \hline \hline
\textbf{AES-256 based}& \cmark  & \cmark   & \xmark       & \cmark        \\
\textbf{SHA-3 based}& \cmark & \cmark      & \cmark       & \cmark        \\ \hline \hline
\end{tabular}
}
\caption{PRNGs used in PQC DS Algorithms.}
\label{tab:rngUsageinPQCDS}
\end{table}

\subsubsection{\textbf{Lattice-based DS}} Lattice DS implementations use PRNGs to generate secret and error components during key-pair and nonce generation ($y$) during signature generation (\textit{Sign()}) modules~\cite{ravi2018side,primas2017single,ravi2019exploiting}. Prior knowledge of $y$ or information about its repeated usage for different messages compromises the security of the signature~\cite{ducas2018crystals,alkim2019lattice,ravi2019exploiting}. 
We subvert the with PRNG in \textit{Sign()} by replacing AES-256 block cipher in PRNG with a different block cipher \textit{Grasshopper}~\cite{biryukov2015secret}. \textit{Grasshoper} was chosen as the S-box values were not generated pseudo-randomly~\cite{kuznechikBackdoor}. 
\subsubsection{\textbf{Symmetric-based DS}} In SPHINCS+, PRNG is implemented by SHA-based hash using a secret-key $s_k$ and message $M$.
In \textit{sign()}, the PRNG randomly selects a key-pair from the SPHINCS+ tree to sign a message. 
We subvert PRNG by replacing it with a different block cipher (\textit{Grasshopper}~\cite{biryukov2015secret}) based PRNG. The signature generation can leak $s_k$ via less-secure PRNG~\cite{bernstein2019sphincs+}.

\subsubsection{\textbf{Multivariate-based DS}} MQDSS and Rainbow use a PRNG for sampling variable coefficients. These algorithms use secure hashes to implement PRNG using $s_k$ as input. We replace AES in PRNG with \textit{Grasshopper}~\cite{biryukov2015secret} cipher.

\subsection{ASA on Hash functions (HASH)}
Cryptographic hashes SHA-2 and SHA-3 have been used for key-pair generation, signature generation and verification modules across PQC DS.
In table~\ref{tab:hashUsageinPQCDS}, we outline the secure hashes in various PQC DS codes. ASA replaces secure SHA-2/SHA-3 hashes with SHA-0 (Fig.~\ref{fig:pqcanomaly}), which is broken. For compatibility with length of SHA-2/3 outputs, we repeated the sequence of SHA-0's output to match the length.
By using SHA-0 for generating hashes, an adversary can launch  collision attacks~\cite{sha1collision} for key recovery. 
\begin{table}[htbp]
\caption{NIST hashes used by PQC DS Algorithms.}
\begin{center}
\begin{tabular}{lccc|cc}
\hline
\hline
\multirow{3}{*}{\textbf{\begin{tabular}[c]{@{}c@{}}PQC DS\\ Algorithms\end{tabular}}}  & \multicolumn{5}{c}{\textbf{Secure Hash Algorithm}} \\ \cline{2-6} 

& \multicolumn{3}{c|}{\textbf{SHA-2 }} & \multicolumn{2}{c}{\textbf{SHA-3 (SHAKE-)}} \\ \cline{2-6} 
& \textbf{256} & \textbf{384} & \textbf{512} & \textbf{128} & \textbf{256} \\ \hline\hline
Dilithium & \xmark   & \xmark  & \xmark   & \cmark  & \cmark     \\
Falcon     & \xmark & \xmark & \xmark   & \cmark & \cmark   \\
Rainbow  & \cmark   & \cmark    & \cmark    & \xmark & \xmark   \\
SPHINCS+ & \xmark  & \xmark & \xmark & \xmark   & \cmark  \\ \hline
\hline
\end{tabular}

\label{tab:hashUsageinPQCDS}
\end{center}
\end{table}

\subsection{ASA on the Security Parameters (SPARAM)}
 We propose subverting the security parameters for each PQC DS algorithm. The goal of the adversary is to reduce the strength of the generated signature by \textit{modifying crucial Security parameters} in the algorithm. This leads to weaker PQC codes violating NIST security levels. The ASAs the for three classes are:

 \begin{algorithm}[t]
 \small
		\caption{DILITHIUM.Sign($S_{key},\mu$)  \cite{migliore2019masking}}
		\label{algo:dilithiumAnomaly}
            Parameters: {$d,\gamma_{1},\gamma_{2},\beta,\omega,k,l$}
			\begin{enumerate}
			    \item A = PRNG($\rho$) \space   $\in$ \(R_{q}^{k\times l}\)
			    \item $T_{1}$ = Truncate($T,d$) \space  $\in$ \(R_{q}^{k\times 1}\)
			    \item $T_{0}$ = $T$ - $T_{1}$ $\cdot$ $2^d$  \space  $\in$ \(R_{q}^{k\times 1}\)
			    
			    Rejection sampling loop
			    \item $\rho^{"}$ $\leftarrow$ $\{0,1\}^{256}$
			    \item $Y$ = PRNG($\rho^{"}$)  \space  $\in$ \(R_{\gamma_{1}-1}^{l\times 1}\)
			    \item $W$ = $A \cdot Y$ \space $\in$ \(R_{q}^{k\times 1}\)
			    \item $W_{1}$ = $HighBits_{q}(W,2\gamma_{2})$  \space  $\in$ \(R_{q}^{k\times 1}\) 
			    \item $C$ = Hash($\gamma, T_{1},W_{1},\mu$) \space $\in$ $\{0,1\}^{256}$
			    \item $Z$ = $Y + CS_{1}$  \space $\in$ \(R_{q}^{l\times 1}\) 
			    \item $R_{0}$ = $LowBits_{q}(W - CS_{2}, 2\gamma_{2})$
			    \item $H$ = $MakeHint_{q}(-CT_{0},W-CS_{2}+CT_{0},2\gamma_{2})$  
			    
			    \textbf{\bt{Bound Checking}}
			    \item $if \parallel Z \parallel_{\inf}$  $\geq \gamma_{1} - \beta$ goto 4  \textbf{\bt{$\beta= 375$ $\Longrightarrow 0$ \space subverted}}
			    \item $if \parallel R_{0} \parallel_{\inf}$  $\geq \gamma_{2} - \beta$ goto 4  \textbf{\bt{$\beta= 375$ $\Longrightarrow 0$ \space subverted}}
			    \item $if \parallel CT_{0} \parallel_{\inf}$  $\geq \gamma_{2}$ goto 4 
                \item $if H > \omega$  goto 4
			    \item $\sigma$ = $(Z,H,C)$
			    \item return $\sigma$
			    
			\end{enumerate}
	\end{algorithm}

 \subsubsection{\textbf{Lattice-based DS}} In Lattice-based DS, security strength and correctness of a signature is computed by executing bound checks~\cite{alkim2019lattice}. Security parameters control the number of times bound checks are performed. The bounds ensure that the signature does not leak out information about secret key. We subvert the code such that it modifies one of the crucial Security parameters with a lower value. For example, in Dilithium (Algo~\ref{algo:dilithiumAnomaly}), reducing the parameter $\beta$ with a lower threshold ($375 \rightarrow 0$) helps reveal information about the secret-key components ($S_1$, $S_2$) making it vulnerable to forging signatures~\cite{ravi2018side,liusecurity}. SPARAM subversion similarly extends to Falcon as shown in table~\ref{tab:sparamtable}.

\subsubsection{\textbf{Symmetric-based DS}} 
SPHINCS+~\cite{bernstein2019sphincs+} uses Merkle hyper-tree of height $h$, $k$ sub-trees each with $t$ leaves in Forest of Random Subsets ($FORS$). 
The security strength of SPHINCS+ is affected by $h$, $k$ and $t$ \cite{bernstein2019sphincs+}. The classical security of SPHINCS+ \cite{bernstein2019sphincs+} is defined as:
 \begin{equation*}
 \scriptsize
 \begin{aligned}
      b = - \log \left( \frac{1}{2^{8n}} +
       \sum_{\gamma} \left( 1 - \left( 1 - \frac{1}{t} \right)^\gamma \right)^k {q \choose \gamma} \left( 1 - \frac{1}{2^h}\right)^{q-\gamma} \frac{1}{2^{h\gamma}} \right)
\end{aligned}
 \end{equation*}
 
The security parameter is $n$ (is 128 bits for security level I/II), the number of adversarial signature queries to the oracle is $q$ and the bit-level security of SPHINCS+ is $b$. One can weaken 128-bit simple SPHINCS+ by altering the parameters as follows: $h$: $64\rightarrow 8$, $k$: $10 \rightarrow 2$ and $t$: $2^{15} \rightarrow 2^1$. This decreases $b$ violating the security level of SPHINCS+. 
Secret key recovery is easy for the adversary by reducing the number of  queries  $q$ to the oracle. 
 \subsubsection{\textbf{Multivariate-based DS}} Security assumptions in multivariate cryptography are based on the difficulty of solving systems of multivariate polynomials over finite fields (MVP). \textbf{Rainbow}~\cite{rainbow} derives its  structure from \textit{unbalanced Oil and Vinegar} (UOV) scheme. 
To achieve NIST security level I, Rainbow selects the finite-field $\mathcal{F}_q$ with $q$ elements, $m$ multivariate equations and $n$ variables as the security parameters. Here, $m = n - v_1$ and $n=$ $v_1 +o_1 +o_2 $ where $v_1$ is the value of vinegar variable and $o_1$ and $o_2$ are the cardinalities of oil sets $O_1$ and $O_2$ respectively. In order to maintain security level I/II ($seclev$), $m \geq \frac{2.seclev}{\log_2q}$ must be satisfied by setting $m$ to be at least 64. We subvert the code by changing $o_1$: $32 \rightarrow 16$, $o_2$: $32 \rightarrow 16$ resulting in $m=32$. This makes the implementation prone to collision attacks.
\begin{table}[!htb]
\caption{Subverting security parameters (SPARAM) in PQC digital signatures.}
\begin{center}
\begin{tabular}{llll}
\hline \hline
\multirow{2}{*}{\textbf{PQC DS}} & \multicolumn{2}{c}{\textbf{SPARAM (NIST Security level-I)}}  \\ \cline{2-3} 
        & \textbf{Original}   & \textbf{Subverted}  \\ \hline \hline
Dilithium    & $\beta=375$   & $\beta=0$        \\
Falcon       & $\beta =6599$    & $\beta=0$     \\
Rainbow      & $o_1=32,o_2=32$  & $o_1=16,o_2=16$ \\
SPHINCS+     & $h=64,k=10,t=2^{15}$   & $h=8,k=2,t=2^1$     \\ \hline \hline
\end{tabular}

\label{tab:sparamtable}
\end{center}
\end{table}

\section{Limitations of State-of-the-Art Detectors}
\label{sec:limitationsOfSOTA}
In section~\ref{sec:anomalies}, we injected ASAs~\cite{debssl2008,dual_ec_2016} into PQC DS codes to create subverted implementations. We  outline limitations of ML-based HPC detectors in detecting ASA-compromised codes and motivate our approach.
We will study whether ML-based HPC detectors can distinguish a trusted implementation from a subverted one.
\subsection{Preliminary Findings}
\label{sec:findingsSOTA}
\begin{figure*}[t]
    \centering
   \subfloat[random inputs]{\includegraphics[width=0.48\columnwidth]{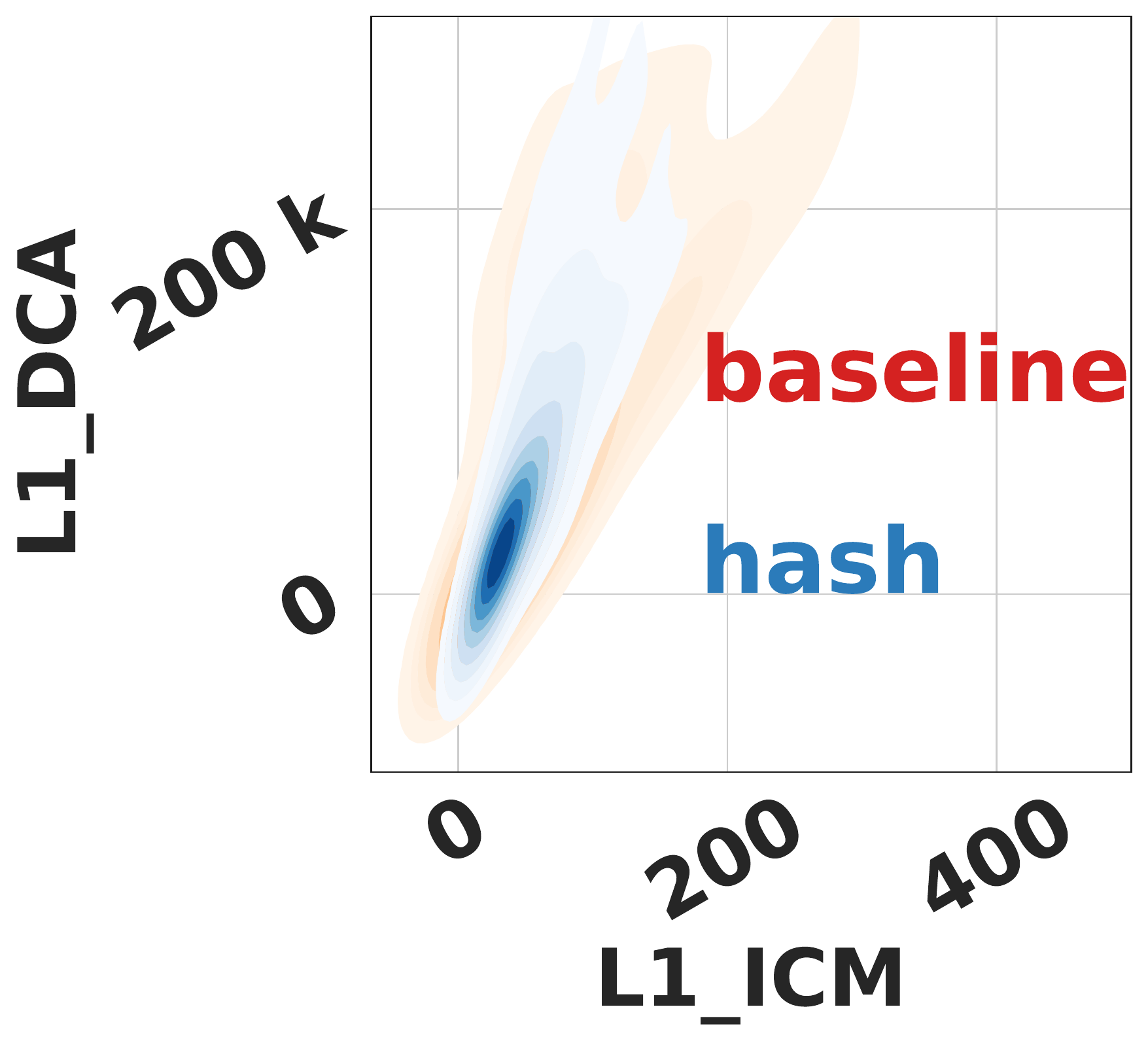}}%
   \subfloat[random inputs]{\includegraphics[width=0.44\columnwidth]{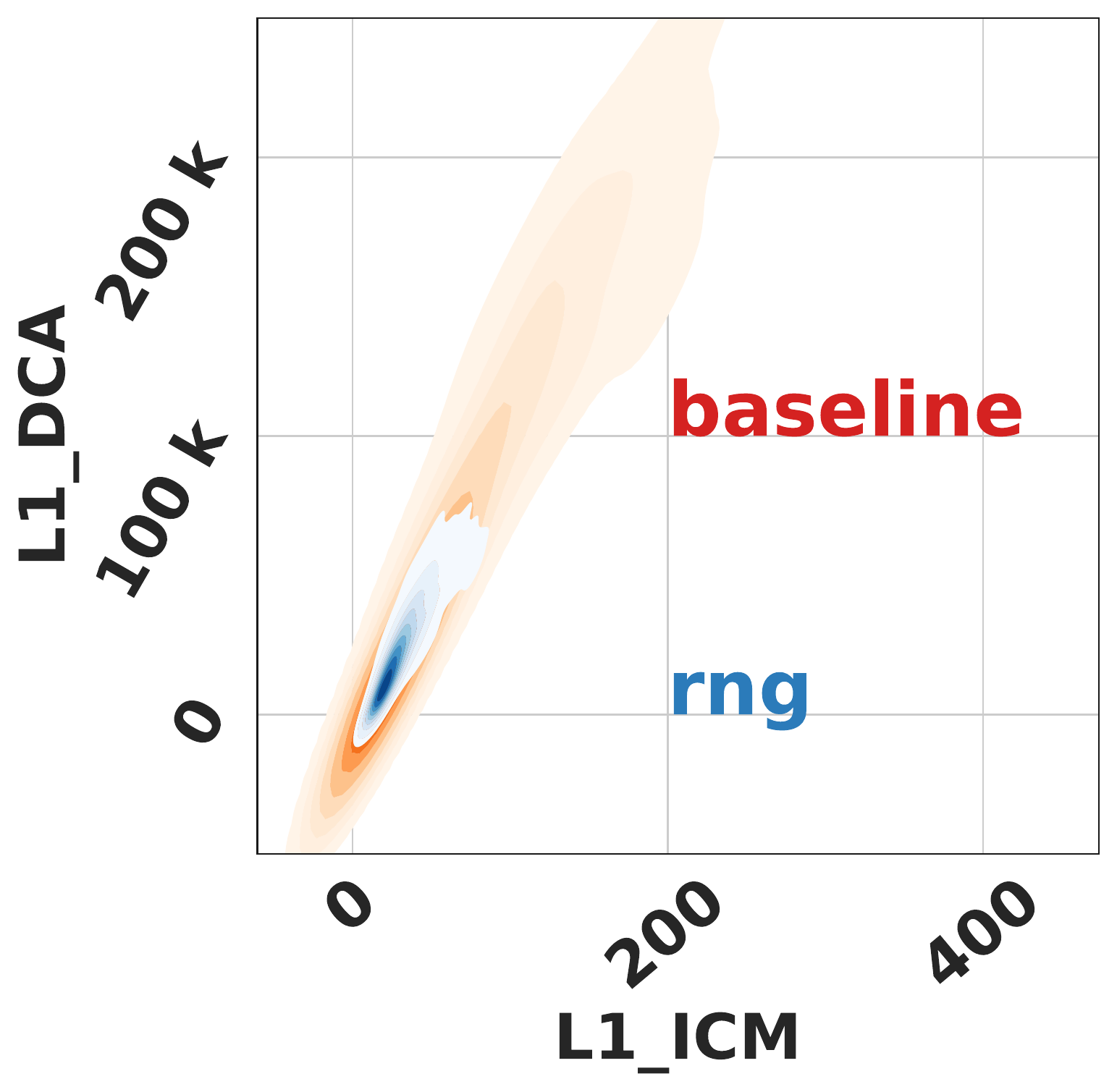}}%
    \subfloat[fuzzed inputs]{{\includegraphics[width=0.48\columnwidth]{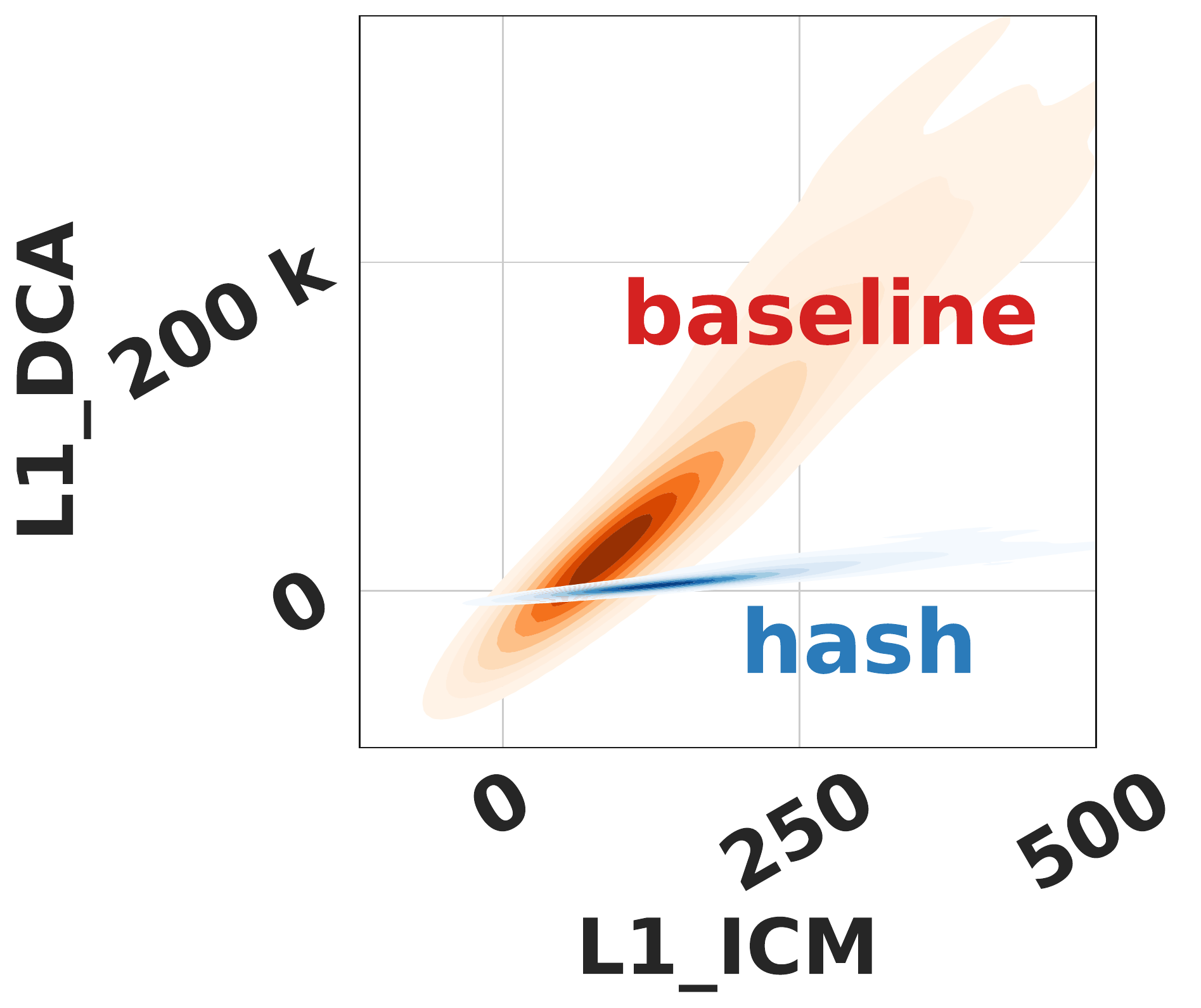}}}%
    \subfloat[fuzzed inputs]{{\includegraphics[width=0.44\columnwidth]{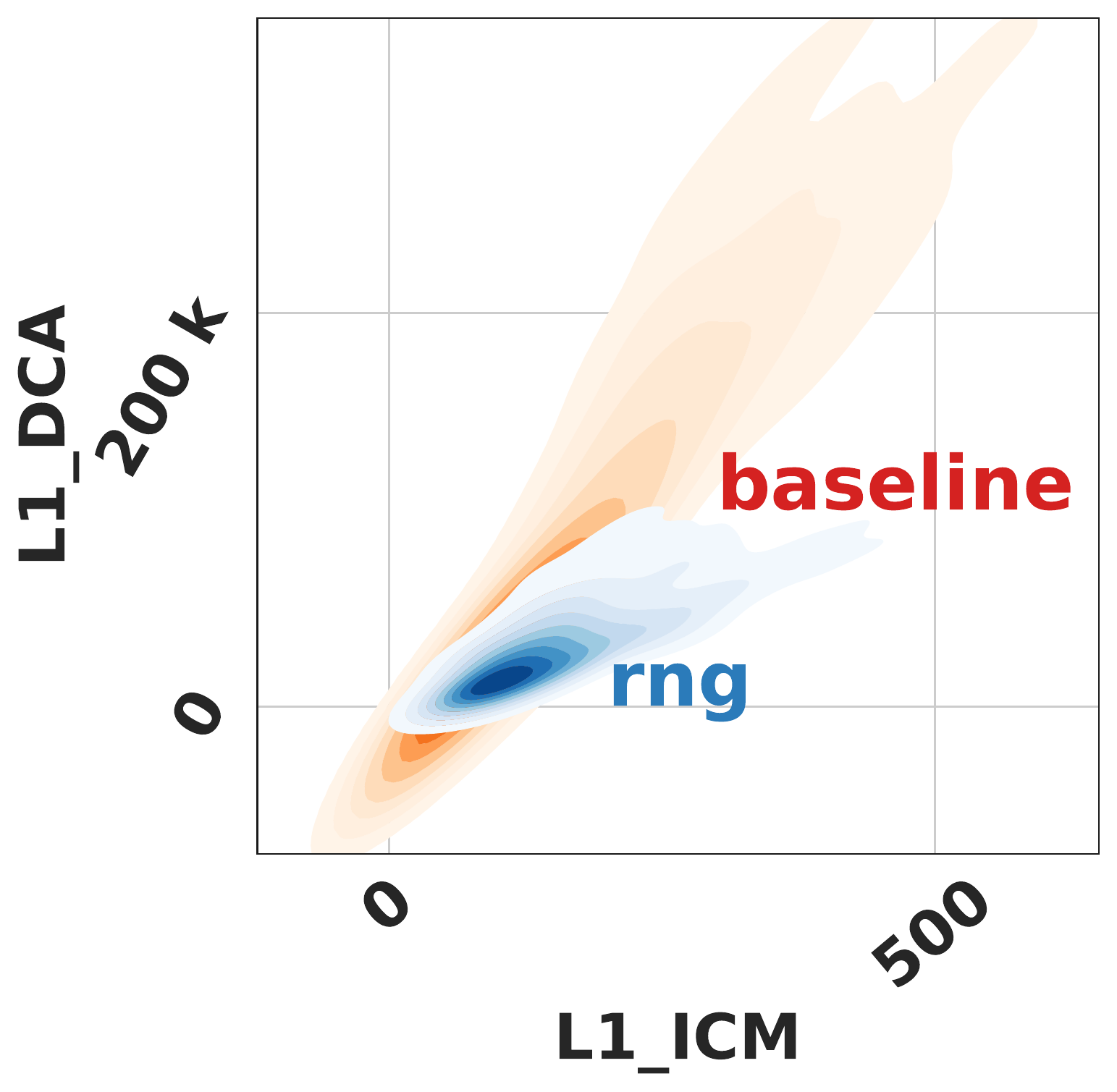}}}%
    \caption{Kernel density estimation plots for HPC pairs at a PC in trusted and (HASH and PRNG) ASA code executions of Dilithium using (a,b) random (c, d) fuzzed inputs.}
    \label{fig:motivFuzz}
\end{figure*}

We collect HPC side-channels and use ML-based classifier to detect traces of ASA-compromised  implementations. Results show that about 95\% of HPC traces of ASA-compromised implementations are similar to that for original implementations. In Fig.~\ref{fig:motivFuzz}a and ~\ref{fig:motivFuzz}b, the spread of HPC values with random inputs overlaps for trusted and ASA implementations. This motivates us to find out the shortcomings of such detection techniques. We analyzed the traces from the profiled code and make two key observations:
\begin{itemize}
    \item Most inputs (messages + secret keys) used to generate digital signatures have poor code coverage. Random inputs do not go deep into the code segments. This led to the indistinguishable HPC traces for trusted and ASA implementations.
    \item PQCs rely a lot on randomness. Given an input (message + key), the output signature differs non-determinstically based on the seed used by the implementation.
\end{itemize}

\subsection{Improving the Coverage}

\begin{figure}[h]
    \centering
    \includegraphics[width=0.6\columnwidth]{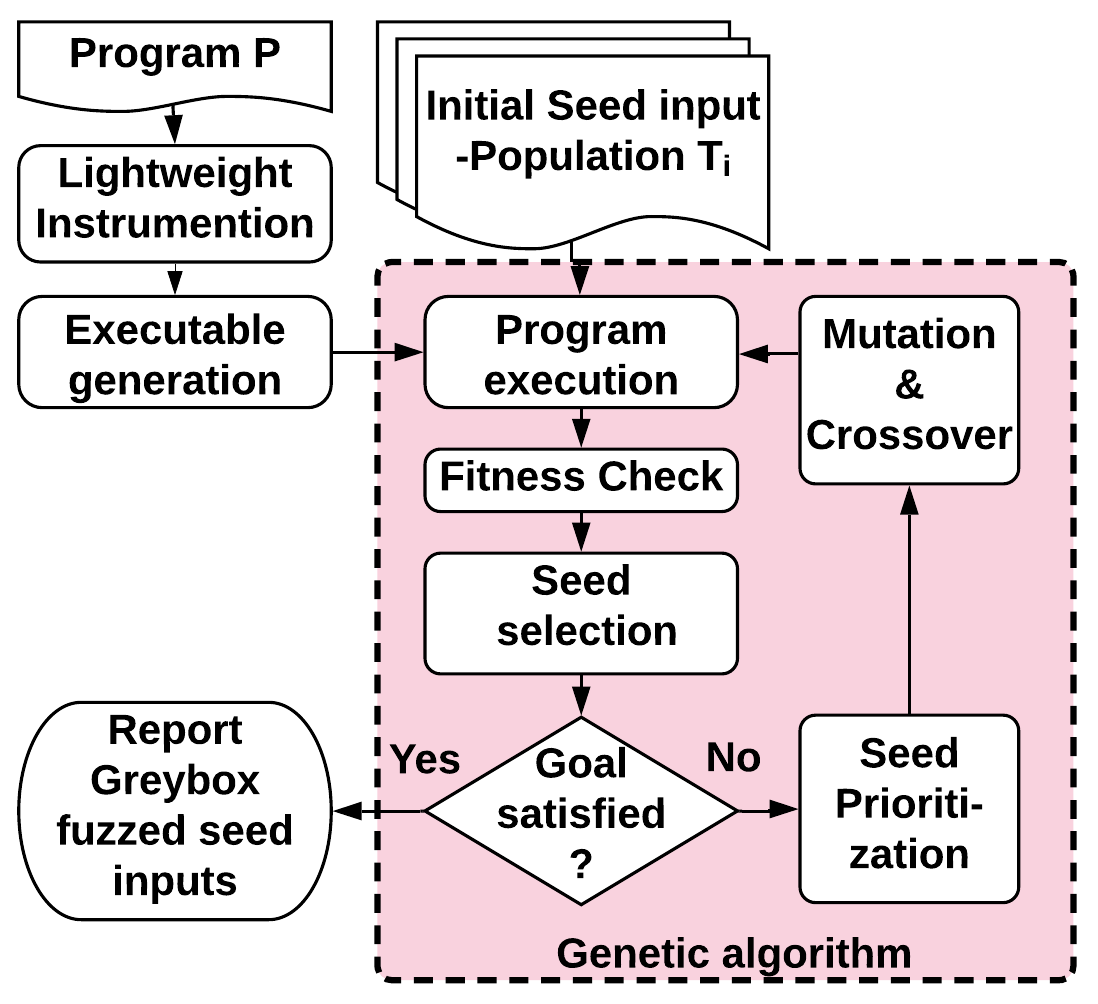}
    \caption{Coverage-based greybox fuzzing.}
    \label{fig:CGF}
\end{figure}

Fuzzing~\cite{fuzzing2007} is an automated test generation technique used to check the robustness of software. Our work leverages coverage-guided Greybox fuzz testing~\cite{afl2014} to generate intelligent seed inputs to maximize edge coverage~\cite{mcdc1994} on the control-flow graph of a PQC code. A control-flow graph (CFG) represents the flow of sequential program statements (basic blocks) based on conditions. The technique annotates every edge of the CFG and uses an evolutionary algorithm with fitness functions to monitor run-time behavior.
Fig~\ref{fig:CGF} represents a typical flow of Greybox fuzzing.


\begin{table}[!htb]
\caption{Coverage achieved using random and Greybox fuzzed seed inputs on PQC DS Algorithms}
\begin{center}
\begin{tabular}{lcccc}
\hline \hline
\multicolumn{1}{c}{\multirow{2}{*}{\textbf{\begin{tabular}[c]{@{}c@{}}Coverage\\ metric\end{tabular}}}} & \multirow{2}{*}{\textbf{\begin{tabular}[c]{@{}c@{}}PQC DS \end{tabular}}} & \multicolumn{3}{c}{\textbf{Coverage achieved}}                                                                                                                                                                            \\ \cline{3-5} 
\multicolumn{1}{c}{}                                                                                            &                                                                                            & \textbf{\begin{tabular}[c]{@{}c@{}}Random\\ Seed \end{tabular}} & \textbf{\begin{tabular}[c]{@{}c@{}}Fuzz\\ Seed\end{tabular}} & \textbf{\begin{tabular}[c]{@{}c@{}}Improve\\ (\%)\end{tabular}} \\ \hline \hline
\multirow{3}{*}{\textbf{\begin{tabular}[c]{@{}l@{}}Basic\\ blocks\\ covered\end{tabular}}}                      & Dilithium & 594 & 687  & 15.65 \\
                   & Falcon & 827 & 870 & 5.19  \\ 
                   & Rainbow & 791 & 815 & 3.03 \\
                   & SPHINCS+ & 1623 & 1812 & 11.64 \\\hline
\multirow{3}{*}{\textbf{\begin{tabular}[c]{@{}l@{}}Control \\ flow edge\\ covered\end{tabular}}}    & Dilithium & 1549 & 2093 & 35.11 \\
                          & Falcon  & 1088 & 1314 & 20.77 \\
                          & Rainbow  & 1563 & 1696 & 8.57 \\ 
                          & SPHINCS+  & 1215 & 1594 &  31.19\\ \hline \hline
\end{tabular}

\label{tab:fuzzCoverage}
\end{center}
\end{table}

Obtaining high quality seed inputs which can provide best coverage is important as subverted codes are stealthy and trigger at certain program states. Prior work~\cite{sage2012} has shown malicious subverting maybe inserted in complex conditional checks. Random seed inputs provide shallow coverage and cannot trigger stealthy subversion. HPC values obtained for a seed input can be directly correlated to its execution profile~\cite{ammons1997exploiting}. 
Execution traces using random seed inputs on trusted and subverted code will be exact, yielding indistinguishable HPC signatures in subverted PQC codes. HPC signature should be collected using inputs that maximize state-space coverage. 
Seed inputs from Greybox fuzzing give statistical guarantees about achievable program state coverage under a specific time constraint~\cite{bohme2018stads}. 
In table~\ref{tab:fuzzCoverage}, we compare basic-block and edge coverage of PQC DS codes using random seeds and seeds generated using Greybox fuzzing. There is upto $\sim$35 \% improvement in edge coverage 
relative to random seeds.  


These findings lead us to propose our core contribution: combine testing with HPC-based detection. We leverage light-weight state-of-art greybox fuzzer AFL, to generate inputs that improve coverage on PQC implementations. Fuzz generated test-inputs have superior distinguishablity power on HPC traces (see Fig.~\ref{fig:motivFuzz}c \& ~\ref{fig:motivFuzz}d).  We use these inputs to extract meaningful features from HPC traces, train our ML-classifier and deploy it to detect a subverted implementation. Our work offers a systematic approach to help NIST generate good quality inputs for detecting vulnerable implementations. In next section, we discuss our detection scheme.

\section{HPC-based Detection of Algorithm Subversion}
\label{sec:methodology}

\begin{figure}[t]
    \centering
\includegraphics[width=\columnwidth]{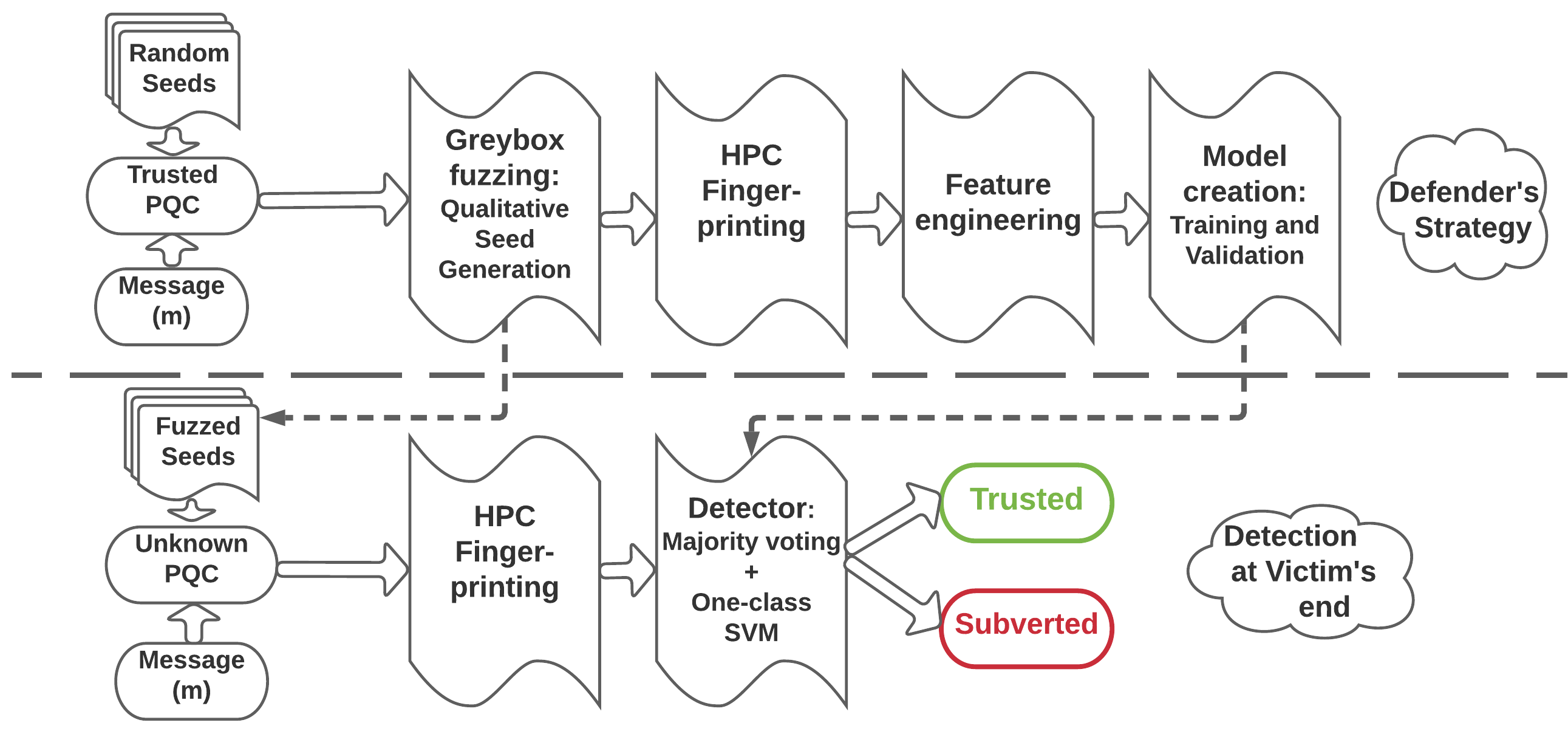}
    \caption{Two-Phase Approach to HPC-based Detection of Algorithm Subversion Attacks.}
    \label{fig:hpcflow}
\end{figure}

We present ML-based methodology to detect subverting of PQC DS codes using HPCs combined with greybox fuzzing as a pre-processing step. 
We use a (i) time-series approach to capture the temporal variations of HPCs and (ii) \textit{program checkpoint (PC)} to capture spatial variations of HPCs in program-flow, as HPC signatures of the PQC codes. 
The PC-based approach monitors the HPC values at every checkpoint in the code. Any modification in the code is expected to cause variation of HPC values during run-time in one (or more) PCs detecting ASA. Fig.~\ref{fig:hpcflow} explains our two-phase approach: The offline phase entails running a trusted PQC code with inputs generated by greybox fuzzing to form HPC signatures. Features from the HPC signatures are trained using ML methods. While detection at end-user(\textit{victim}), given a subverted PQC code from a third party, the trained ML models are deployed to monitor HPC signatures to predict if PQC code was subverted. We describe the steps in this section.





\subsection{HPC Signature Collection}
We use Greybox fuzzing as our first step to generate qualitative seed inputs $S$ for a trusted PQC code $P$. Using $S$, we collect HPC signatures of the \textit{sign()} component of $P$. We present a time-series ($P^{TS}_{sign}$), and PC HPC signature ($P^{PC}_{sign}$) for fingerprinting $P$.

\subsubsection{Time-series based HPC Signature Collection} We run $P$ using a seed input $s \in S$ for a time-period $T_M$. HPC signatures are collected at sampling interval of $T_S$ with $N$ samples, each having $K$ HPC values. This HPC signature obtained across the monitored time-period fingerprints the PQC code.
Algorithm~\ref{algo:timeSeriesHPC} summarises this approach. 

\begin{algorithm}[!htb]
\SetKwInput{KwInput}{Input}                
\SetKwInput{KwOutput}{Output} 
{\small
 \caption{Time-series HPC Signature Collection}
 \KwInput{DS Program $P$,  Monitored set of HPCs : $\{HPC_{idx} | idx \in [1,K] \}$, Sampling interval : $T_S$, Monitored time-period : $T_M$} 
 \KwOutput{Time-series HPC Signature: $P^{TS}_{sign} \in \mathbb{R}^{N \times K}$, 
 }
 
 $P^{TS}_{sign} \leftarrow \phi$\;
 Number of samples collected ($N$)  $\leftarrow \floor{T_M/T_S}+1$ \;
 \For{$idx\gets 1$ \KwTo $K$}{
  \For{$i \gets 1$ \KwTo $N$}{
        Collect $HPC^i_{idx}$.
  }
    
   }
   $P^{TS}_{sign} \leftarrow HPC^{N}_{K}$\;
   
   
  \Return $P^{TS}_{sign}$.
  \label{algo:timeSeriesHPC}
  }
\end{algorithm}

\subsubsection{PC-based Signature Collection} We insert PCs and run $P$ using seed inputs $S$ from fuzzing. HPC signatures are collected across multiple PCs $H$ as shown in Algorithm~\ref{algo:programCheckpointHPC}. 

\begin{algorithm}[!htb]
\SetKwInput{KwInput}{Input}                
\SetKwInput{KwOutput}{Output} 
{\small
 \caption{Program Checkpoint (PC) HPC Signature}
 \KwInput{DS Program $P$,  Monitored HPCs: $\{HPC_{idx} | idx \in [1,K] \}$ , Seed inputs: $S$, Checkpoints: $H$} 
 \KwOutput{PC-based HPC Signature: $P^{PC}_{sign}  \in \mathbb{R}^{(S * H )\times K}$ }
 
 $P^{PC}_{sign} \leftarrow \phi$\;
 \For{seed input $i \in S$}{
    \For{PC $j \in H$}{
        Collect $HPC^{ij}_{idx}$, $\forall idx \in [1,K]$
    }
   }
 $P^{PC}_{sign} \leftarrow HPC^{S * H}_{K}$\;
 
  \Return $P^{PC}_{sign}$
  \label{algo:programCheckpointHPC}
  }
\end{algorithm}

\subsection{HPC Selection} 
The ARM Cortex A8 processor has 16 HPCs. However, the processor restricts use of four HPCs simultaneously. All HPCs can be monitored by time-multiplexing. It is crucial to select the set of HPCs that can uniquely characterize the program behavior.
We use metrics from \cite{tang2014unsupervised}  to identify  HPCs that can uniquely fingerprint the codes: 
\begin{itemize}
\item  Principal Component Analysis (PCA): Select HPCs with maximum variance by creating uncorrelated variables from orthogonal components that maximise variance~\cite{chetsa2014exploiting}.
\item Maximum Standard Deviation threshold: Identify HPCs with high deviation from mean exceeds a threshold.
\item Maximum Variance threshold: Identify HPCs with high variance whose variance exceeds a threshold value.
\item Fisher Score (F-score): Measures discriminative power of HPCs that have large separation between their deviation and mean for data of different classes.
\end{itemize}
We use PCA to identify  HPCs with maximum variance.
\subsection{Feature Engineering}
Feature engineering is an key aspect of our  methodology. Our goal is to generate features from the HPC signature to train our ML-classifiers for every PQC algorithm.

\subsubsection{Time-series based}
To extract features, we apply \textit{overlapping sliding time-window} and generate windowed sub-sequence of time-series data. We define two parameters : time-window duration  ($T_{len}$) and time-window shift ($T_{shift}$). $T_{len}$ denotes the time-duration considered in the frame of one window and $T_{shift}$ denotes the shift in time-interval between successive time-windows. 
$T_{len}$ and $T_{shift}$ are configurable parameters based on temporal granularity required.
We obtain $D$ windowed sub-sequences. For feature generation, we use the set of HPCs $Z$ obtained from HPC selection.
We extract  mean, maximum, kurtosis, and Kendall-tau correlation coefficient as features from every windowed sub-sequence. Algorithm~\ref{algo:featureEnggtimeSeries} summarizes feature generation $F^{TS}_{sign}$ of dimension  [${D \times 4Z}$] for time-series data.

\begin{algorithm}[t]
\SetKwInput{KwInput}{Input}                
\SetKwInput{KwOutput}{Output} 
{\small
 \caption{Feature Engg. : Time-series}
 \KwInput{Time-series HPC : $P^{TS}_{sign} \in \mathbb{R}^{N \times K} $, Time-window duration : $T_{len}$, Time-window shift : $T_{shift}$, Selected HPCs: $Z$} 
 \KwOutput{Feature Matrix : $F^{TS}_{hpc} \in \mathbb{R}^{D \times 4Z}$}
 \SetKwFunction{FComputeSWFeature}{ComputeSWFeature}
 
 \SetKwProg{Fn}{Function}{:}{}
  \Fn{\FComputeSWFeature{$HPC^N_{K}$,$i$}}{
    $t_{start} \leftarrow (i-1) * T_{shift}$ \;
    $t_{end} \leftarrow t_{start} + T_{len}$ \;
    $W \subset N$ \& $ W \in [t_{start},t_{end})$\;
    $\overrightarrow{\mu_{i}}$ = \textit{Mean}(\{$HPC^j_{l}, j \in W, l \in Z $\})\;
    $\overrightarrow{\kappa_{i}}$ = \textit{Kurtosis}(\{$HPC^j_{l},  j \in W, l \in Z$\})\;
    $\overrightarrow{\tau_{i}}$ = \textit{Corr.}(\{$HPC^j_{l},  j \in W, l \in Z$\})\;
        $\overrightarrow{max_{i}}$ = $\max$(\{$HPC^j_{l},  j \in W, l \in Z$\})\;
        \KwRet [$\overrightarrow{\mu_{i}}$ ,$\overrightarrow{\sigma_{i}}$, $\overrightarrow{\tau_{i}}$,$\overrightarrow{max_{i}}$]\;
  }
 $F^{TS}_{hpc} \leftarrow \phi$\;
 $HPC^{N}_{K} \equiv P^{TS}_{sign}$ (Algo-2)\;
 Number of time-window segments ($D$)  $\leftarrow \floor{T_{M}/T_{shift}} + 1$ \

\For{ $i \gets 1$ \KwTo $D$}{
$F^{TS}_{hpc} \leftarrow F^{TS}_{hpc} \bigcup$ ComputeSWFeature($HPC^{N}_{K}$,$i$) \;
   }
   
  \Return $F^{TS}_{hpc}$.
  \label{algo:featureEnggtimeSeries}
  }
\end{algorithm}

\subsubsection{PC based}
Once HPCs $Z$ are selected, we create a feature matrix $F^{PC}_{hpc}$ of dimension [${(S \times H) \times Z}$] from HPC signature $P^{PC}_{sign}$. Algorithm~\ref{algo:featureEnggPC}  describes how to generate feature matrix for PCs.

\begin{algorithm}[h]
\SetKwInput{KwInput}{Input}                
\SetKwInput{KwOutput}{Output} 
{\small
 \caption{Feature Engineering: Checkpoint  }
 \KwInput{Checkpoint HPC data : $P^{PC}_{sign} \in \mathbb{R}^{(S * H) \times K}$, Selected HPCs: $Z$} 
 \KwOutput{Feature Matrix : $F^{PC}_{hpc} \in \mathbb{R}^{(S * H) \times Z}$}
 
 $F^{PC}_{sign} \leftarrow \phi$\;
 $HPC^{S,H}_{K} \equiv P^{PC}_{sign}$\;
 $S \rightarrow$ Seed inputs obtained to from fuzzing (Algo-3)\;
 $H \rightarrow$ Checkpoints in Program $P$ (Algo-3)\;
 $K \rightarrow$ \# Number of monitored HPCs (Algo-3)\;
\For{seed input $i \in S$}{
    \For{checkpoint $j \in H$}{
        $F^{PC}_{hpc} \leftarrow F^{PC}_{hpc} \bigcup$ $HPC^{i,j}_{Z}$ \;
    }
   }
  \Return $F^{PC}_{hpc}$.
  \label{algo:featureEnggPC}
  }
\end{algorithm}

\subsection{ML Model: Training and Validation}

Using the extracted features, we train the ML models to predict if a PQC code is trusted or subverted. We train two models for every PQC algorithm, using time-series and PC-based features.
Our goal is to train a model that learns trusted data distribution in an unsupervised manner as a one-class problem. Once the model is trained, it predicts if a given test point belongs to the trusted distribution or not.
We use the one-class SVM~\cite{scholkopf2001learning} to map the unlabelled data onto the features space using kernel functions and find the maximum distance of the data bounded by a hyper-plane from the origin. We use the Radial Basis Function (RBF) to  model non-linear representations. 
Training the model with RBF requires parameter tuning to generalize the model. The model uses parameters $\gamma$ and $\nu$. Increasing $\gamma$ decreases regularization value and finds the optimal hyper-plane and $\nu \in (0,1]$  controls the trade-off between penalty of mis-classification and generalization of parameters. We tune $\gamma$ and $\nu$ separately for time-series and PC-based features.
\subsubsection{Time-series} 
We use time-series features to train a one-class SVM model for every PQC algorithm ($Model_{TS}$). We use 90\% of  trusted feature set for training and rest for testing. During training, we use feature set $F^{TS}_{sign}$ to train each model with its own hyper-parameters.
We use three steps to evaluate model performance on test dataset.
\begin{enumerate}
\item Given a feature set of $X_{ts}$ observation vectors (i.e., feature vectors of $X_{ts}$ time-windows), on prediction, they yield trusted (1) or subverted (-1) labels.
\item We temporally aggregate prediction label set ($P_{L}$). We define a majority threshold as $t_{ts}\in X_{ts}$, divide into subsets, where each subset has $t_{ts}$ prediction labels. The number of subsets, $N_{P}$=$\floor{X_{ts}/t_{ts}}$. In every subset, we aggregate labels as: $\sum_{i=1}^{t_{ts}} P_{i}$, $P_{i}\in P_{L}$. If this sum is positive, the prediction label for that subset is assigned as trusted else subverted. 
\item We compute accuracy: (\# correctly predicted subsets) / (\# subsets).
\end{enumerate}
We use temporal aggregation with majority threshold to reduce mis-classification rates. Temporal aggregation enables mitigating errors caused by mis-predictions of certain time-windows. The best threshold is selected via experimentation.

\begin{figure*}[t]
    \centering
    \subfloat{{\includegraphics[width=1.7\columnwidth]{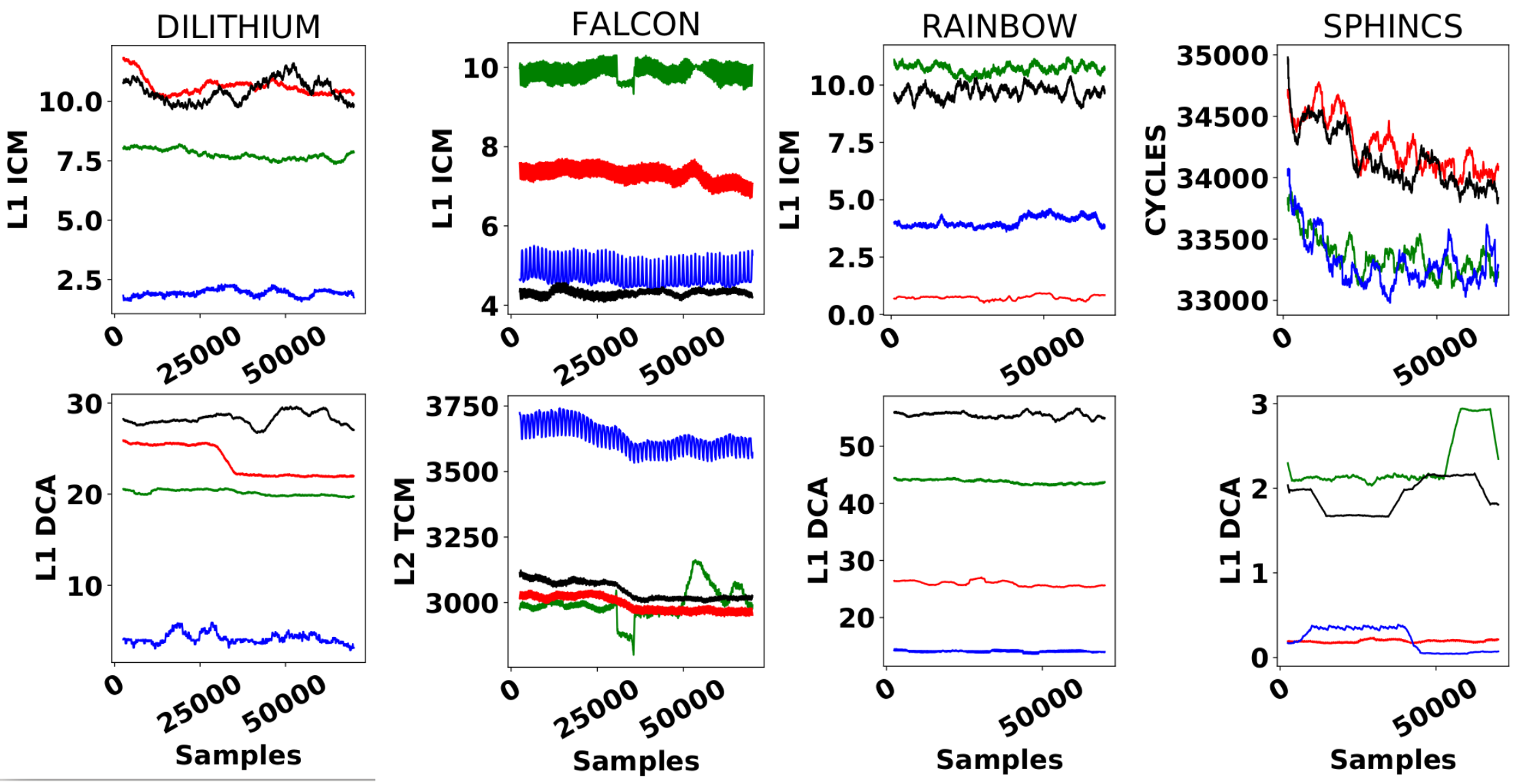}}}%
    \caption{Time-series HPC variations of Trusted (Green) and subverted codes (SPARAM (Red), PRNG (Blue), and HASH (Black)). HPCs are selected and monitored for four PQCs.}
    \label{fig:tsplots}
\end{figure*}

\subsubsection{Program Checkpoints (PC)}
We use PC as a feature of the trusted PQC codes, $F^{PC}_{hpc}$ and develop ensembles of one-class SVM models ($Model_{PC}$). Our ensemble models are formed by dividing HPC features into smaller subsets, and train one-class SVM models on each feature subset. $Model_{PC}$ = $Ensemble\{SVM(F^{PC}_{hpc_{1-4}})$, $SVM(F^{PC}_{hpc_{5-8}})\}$. We train models on smaller subset of HPCs so that they learn different behavioral characteristics. 
We use 66.6\% of the baseline dataset for training.
We follow four steps to evaluate performance on any test dataset.
\begin{enumerate}
    \item Given a feature matrix containing $X_{pc}$ observations, on prediction yields prediction labels ($P_{h}$) of length $X_{pc}$, as 1 or -1, i.e., trusted or subverted respectively.
    \item We perform spatial aggregation on the prediction labels. We define a majority threshold, $t_{pc}\leq X_{pc}$, where the set is divided into smaller subsets, such that each subset has $t_{pc}$ labels. Thus, the number of subsets becomes $N_{h}$ = $\floor{X_{pc}/t_{pc}}$. For every subset, we aggregate the labels as: $\sum_{i=1}^{t_{pc}} P_{i}$, $P_{i}\in P_{h}$. If the sum is positive, the prediction label for that subset is assigned trusted else subverted obtaining a set of prediction labels of size $N_{h}$.
    \item We compute accuracy as: (number of truly predicted subsets) / (number of subsets).
    \item We perform three-fold cross validation on the trusted dataset. We select the seed inputs ($Y_{s}$) from the set which performs best during cross validation.
\end{enumerate}

\subsection{Detection} 
 The victim perform spatial and temporal checking by deploying models, $Model_{TS}$ and $Model_{PC}$ in his environment. The victim collect PC-based and time-series based HPC signatures from the untrusted PQC codes using $S$ seed inputs and the given message. While $Model_{TS}$ performs dynamic run-time detection continuously, $Model_{PC}$ performs a one-time check by predicting observations using $Y_{s}$ seed inputs.

\section{Experimental Results}
\label{sec:expresults}
\subsection{Experimental Setup}
We performed our experiments on NIST Round 3 digital signature algorithms - Dilithium, Falcon, Rainbow and SPHINCS+ from the alternative candidate. The experiments were run on 32-bit ARM Cortex A8 processor running Linux kernel 4.14 with clock-frequency of 1.5 GHz. We report detailed breakdown of time consumed at each phase of our proposed HPC-based detection in ~\autoref{tab:timingComplexity}.
\subsection{HPC Variations}
\begin{figure*}[!htb]
    \centering
   \subfloat{\includegraphics[width=2\columnwidth]{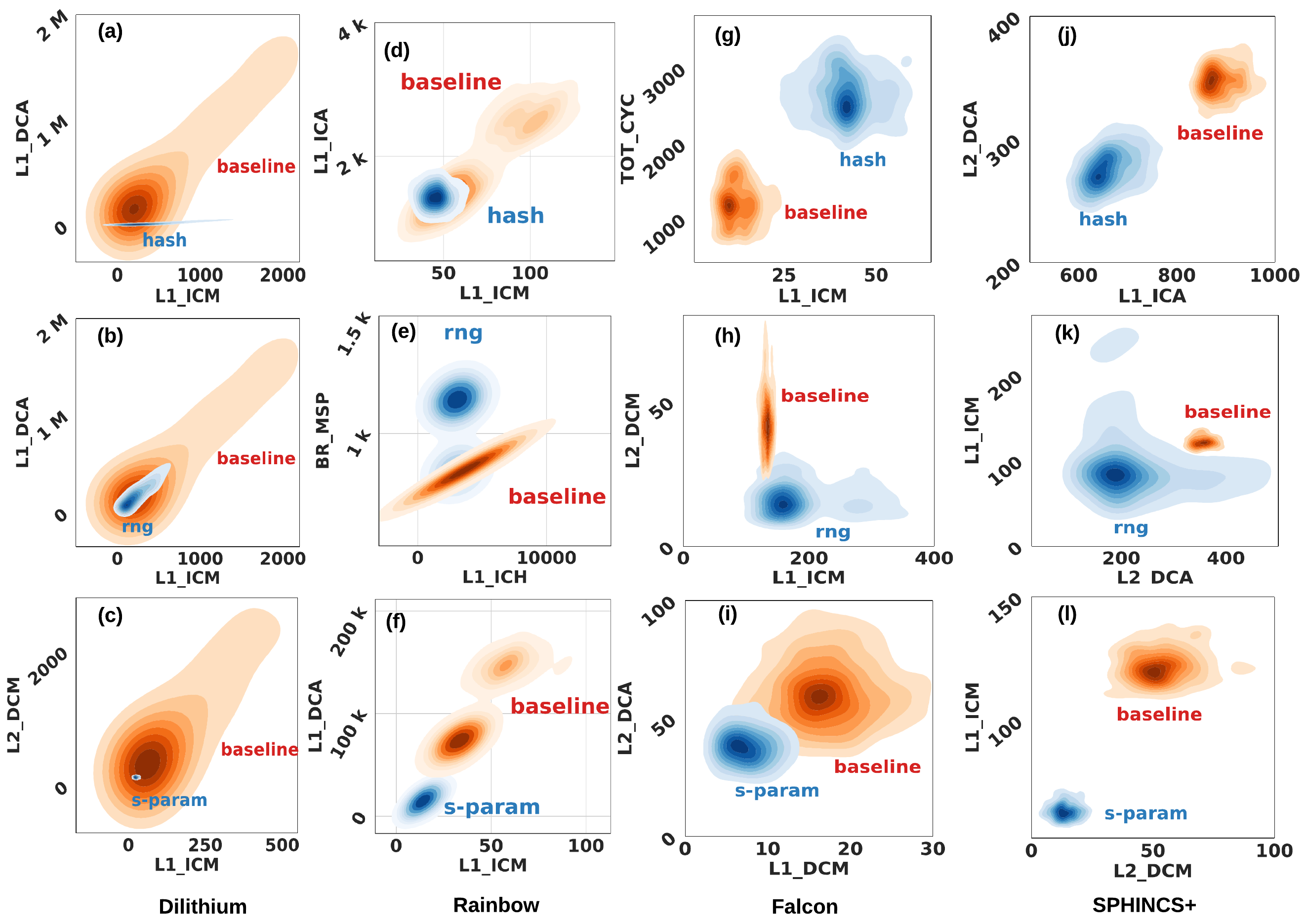}}%
    \caption{Kernel density estimation plots show correlation between pairs of HPCs for program checkpoints (PCs) in trusted and subverted (HASH, PRNG, SPARAM) PQC code executions. Dilithium: (a) PC9 (b) PC9 (c) PC10. Rainbow:  (d) PC5 (e) PC2 (f) PC9. Falcon:  (g) PC35 (h) PC31 (i) PC18. SPHINCS+: (j) PC1 (k) PC4 (i) PC7. }
    \label{fig:pckde}
\end{figure*}

We collect time-series $P^{TS}_{sign}$ and program checkpoint $P^{PC}_{sign}$ HPC signatures for six PQC DS codes. We visualize the variations observed in time-series and program checkpoint HPC signatures in figures \ref{fig:tsplots} and \ref{fig:pckde}.

\subsubsection{Time-series} We run the \textit{Sign()} module of a PQC DS code in an infinite loop at a sampling frequency of 100KHz for 1s, to collect $P^{TS}_{sign}$. We run a lightweight time-series HPC measurement tool parallel to collection of HPCs. We obtain time-series features ($F^{TS}_{hpc}$) using a time-window of 1000 samples with an overlap of 100 samples between successive windows. We set $T_{shift}$ = 0.001s and $T_{len}$ = 0.01s.  We execute trusted and subverted code (PRNG, HASH, SPARAM) variants for considered PQC algorithms and collect the HPC signatures. Fig.~\ref{fig:tsplots} show variations of HPC signatures between trusted and subverted PQC executions. HPCs are selected separately for each trusted PQC code using principal component analysis (PCA) creating trusted HPC signature. For Dilithium, L1-ICM and L1-DCA are selected, and Fig.~\ref{fig:tsplots} show that they discriminate the signature of the trusted implementation from subverted implementation. 
In Falcon, while trusted code executions incur large L1-ICM, the subverted implementations yield lower L1-ICM. So L1-ICM has high distinguishability power for Falcon.
The plots also show that L2-TCM for Falcon and L1-DCA for SPHINCS+ cannot distinguish trusted from subverted implementations. 

\subsubsection{Program Checkpoints (PCs)}
We execute trusted and subverted PQC implementations with seed inputs obtained by fuzzing. We generate program checkpoint (PC) based HPC signatures. We show the Gaussian Kernel Density Estimation (KDE) of pairs of HPCs across a  set of checkpoints (Fig.~\ref{fig:pckde}). To reiterate, choice of HPCs is guided by PCA. We visualize checkpoints where maximum HPC variations are seen.
The plots estimate the probability density of HPC pairs. This aids kernel selection. For SPHINCS+, KDE plots for program checkpoints $pc_1$, $pc_4$ and $pc_7$ (Fig.~\ref{fig:pckde}j,~\ref{fig:pckde}k,~\ref{fig:pckde}l) indicate that data distribution of L1-ICM, L2-DCM, L2-DCA at those checkpoints are separable by a polynomial kernel. For Dilithium (Fig.~\ref{fig:pckde}a,~\ref{fig:pckde}b,~\ref{fig:pckde}c), HPC data-distribution of subverted code executions lies within that of the trusted code execution. Therefore, an RBF kernel can learn the separability of the data for training the SVM.
These KDE plots aid us select kernel function of one-class SVM for training our model. Empirical evaluations deduce that for an RBF kernel $\gamma$=0.01-0.0001 and $\mu$ = 0.1-0.4 are parameters for favourable model performance.   

\begin{table}[t]
\caption{Timing complexity of proposed detection}
\resizebox{\columnwidth}{!}{
\begin{tabular}{llrrrr}
\toprule
\multicolumn{2}{l}{\textbf{\begin{tabular}[c]{@{}l@{}}PQC Signatures $\rightarrow$\\ Time Taken(in s) $\downarrow$\end{tabular}}}                                                  & \multicolumn{1}{l}{\textbf{Dilithium}} & \multicolumn{1}{l}{\textbf{Falcon}} & \multicolumn{1}{l}{\textbf{SPHINCS+}} & \multicolumn{1}{l}{\textbf{Rainbow}} \\ \midrule
\multirow{3}{*}{Offline}                                                           & \begin{tabular}[c]{@{}l@{}}Fuzz seed \\ generation\end{tabular} & 18000                                  & 18000                               & 18000                                 & 18000                                \\
                                                                                   & HPC extraction                                                  & 2.958                                  & 20.468                               & 209.210                                & 2.112                                \\
                                                                                   & Training                                                        & 4.58                                   & 4.64                                & 4.54                                  & 4.38                                 \\ \midrule
\multirow{2}{*}{\begin{tabular}[c]{@{}l@{}}Online\\ (TS)\end{tabular}} & HPC extraction                                                  & 10                                     & 20                                  & 10                                    & 20                                   \\
                                                                                   & Detection                                                       & 3.7                                    & 3.6                                 & 3.6                                   & 3.7                                  \\ \midrule
\multirow{2}{*}{\begin{tabular}[c]{@{}l@{}}Online\\ (PC)\end{tabular}} & HPC extraction                                                  & 0.687                                  & 4.68                                & 48.4                                  & 0.499                                \\
                                                                                   & Detection                                                       & 1.2                                    & 1.35                                & 0.8                                   & 1.34                                 \\ \bottomrule
\end{tabular}
}
\label{tab:timingComplexity}
\end{table}

\begin{table}[t]
\begin{center}
\caption{HPC time-series for different thresholds ($t$) over time windows: Positive ($Pos$), Negative ($Neg$) predictions for trusted and SPARAM, PRNG, HASH subverted codes.}
\resizebox{\columnwidth}{!}{
\begin{tabular}{cccccccc}
\hline \hline
\textbf{}          & \textbf{}     & \multicolumn{6}{c}{\textbf{\begin{tabular}[c]{@{}c@{}}Predicted Type\end{tabular}}}                                                    \\ \cline{3-8} 
\textbf{Algorithm} & \textbf{Variant} & \multicolumn{2}{c}{$\mathbf{t_{ts}=21}$} & \multicolumn{2}{c}{$\mathbf{t_{ts}=31}$} & \multicolumn{2}{c}{$\mathbf{t_{ts}=41}$} \\ \cline{3-8} 
                   &  \textbf{Type}  & $\mathbf{Pos}$        & $\mathbf{Neg}$       &$\mathbf{Pos}$        & $\mathbf{Neg}$ & $\mathbf{Pos}$        & $\mathbf{Neg}$    \\ \hline \hline
Dilithium          & Trusted      & 1.0       & 0.0      & 1.0        & 0.0      & 1.0      & 0.0      \\ 
                   & SPARAM          & 0.02    & 0.98     & 0.02  & 0.98   & 0.02 & 0.98 \\ 
                   & PRNG     & 0.0    & 1.0   & 0  & 1.0    & 0  & 1.0      \\ 
                   & HASH          & 0     & 1.0      & 0    & 1.0     & 0   & 1.0      \\\hline
Falcon              & Trusted    & 1.0 & 0.0   & 1.0       & 0.0   
& 1.0   & 0.0   \\
                    & SPARAM      & 0.0   & 1.0   & 0.0   & 1.0   & 0.0   & 1.0
                    \\
                    & PRNG  & 0.0   & 1.0   & 0.0   & 1.0   & 0.0   & 1.0
                    \\
                    & HASH       & 0.0   & 1.0   & 0.0   & 1.0   & 0.0   & 1.0
                    \\ \hline
SPHINCS+             & Trusted  & 0.79  & 0.29  & 0.71  & 0.29  & 0.63  & 0.37 \\
                    & SPARAM      & 0.0   & 1.0   & 0.0   & 1.0   & 0.0   & 1.0
                    \\
                    & PRNG & 0.28  & 0.72  & 0.26 & 0.74   &   0.27    & 0.73 \\
                    & HASH   & 0.21  &  0.79 & 0.22  & 0.78  &   0.21    & 0.79 \\ \hline
Rainbow             & Trusted  & 1.0 & 0.0   & 1.0       & 0.0   
& 1.0   & 0.0   \\
                    & SPARAM      & 0.0   & 1.0   & 0.0   & 1.0   & 0.0   & 1.0
                    \\
                    & PRNG & 0.0   & 1.0   & 0.0   & 1.0   & 0.0   & 1.0
                    \\
                    & HASH   & 0.0   & 1.0   & 0.0   & 1.0   & 0.0   & 1.0
                    \\ \hline \hline

\end{tabular}
}

\label{tab:tsModel}
\end{center}
\end{table}

\subsection{Code Subversion Detection Results}
Subversion detection using $Model_{TS}$ and $Model_{PC}$ are shown in Tables \ref{tab:tsModel} and \ref{tab:pcModel}. We evaluate using trusted/subverted PRNG, HASH, SPARAM HPC datasets. In Tables~\ref{tab:tsModel} and \ref{tab:pcModel}, \textit{predicted type} column indicates fractions of test data points predicted as positive ($Pos$) or negative ($Neg$). \textit{Variant type} column lists the type of code subversion. 

\subsubsection{Time-series} In Table \ref{tab:tsModel}, $t_{ts}$ is the majority threshold varied over 21 to 41 time windows. 
Our models for Falcon and Rainbow can detect the trusted code and three subverted codes with 100\% accuracy. There is good temporal granularity amongst the successive time-windows in our data and the methodology mitigates mis-classification rates caused due to a few time-window segments. In SPHINCS+, the accuracy of our model is less than in other algorithms. Since the time-series HPC variations of the trusted code is closer to that of subverted codes, the hyper-plane of the model was not able to distinguish the trusted code from the subverted ones. The subverted PQC codes are so stealthy that their behavior closely matches with the trusted code.
From the Table, we select $t_{ts}$=41 to use in the detection phase for authenticating the integrity of PQC implementation.

\subsubsection{Program Checkpoints} In Table \ref{tab:pcModel}, $t_{pc}$ is the majority threshold varied from 11 to 31 seed inputs. 
The Table shows that the PC-based models can identify trusted PQC codes and three subverted code data sets with 100\% accuracy across all $t_{pc}$ values for Dilithium and Falcon. In Rainbow, the performance of the model in identifying  trusted and HASH subverted codes reduces for $t_{pc}$=11 and 21 seeds. The HASH subverted code is hidden in a way that these seeds capture baseline-like behavior. Hence,  HPCs do no show a distinguishable variation. SPHINCS+ plots in Fig \ref{fig:pckde} show that the mean BR-MSP and L1-ICA variations of the HASH subversion is similar to that of the trusted one. $t_{pc}$=31 is the best choice for use in the online phase for one-time verification. Our models can detect a subverted or trusted PQC code with $\sim$100\% accuracy and $\sim$2\% mis-classification.
\begin{table}[t]
\caption{Program checkpoints (PCs) with Majority thresholds (t): Positive ($Pos$), negative ($Neg$) predictions for trusted and SPARAM, PRNG, HASH subverted codes.}
\begin{center}
\resizebox{\columnwidth}{!}{
\begin{tabular}{cccccccc}
\hline \hline
\textbf{}          & \textbf{}     & \multicolumn{6}{c}{\textbf{\begin{tabular}[c]{@{}c@{}}Predicted Type\end{tabular}}}                                                    \\ \cline{3-8} 
\textbf{Algorithm} & \textbf{Variant} & \multicolumn{2}{c}{$\mathbf{t_{pc}=11}$} & \multicolumn{2}{c}{$\mathbf{t_{pc}=21}$} & \multicolumn{2}{c}{$\mathbf{t_{pc}=31}$} \\ \cline{3-8} 
                   &  \textbf{Type}  & $\mathbf{Pos}$        & $\mathbf{Neg}$       &$\mathbf{Pos}$        & $\mathbf{Neg}$ & $\mathbf{Pos}$        & $\mathbf{Neg}$    \\ \hline \hline
Dilithium                  & Trusted                                                                     & 1.0         & 0.0         & 1.0         & 0.0         & 1.0         & 0.0         \\  
                           & SPARAM                                                                         & 0.0         & 1.0         & 0.0         & 1.0         & 0.0         & 1.0         \\ 
                           & PRNG                                                                    & 0.0         & 1.0         & 0.0         & 1.0         & 0.0         & 1.0         \\  
                           & HASH                                                                          & 0.0         & 1.0         & 0.0         & 1.0         & 0.0         & 1.0         \\ \hline
Falcon                     & Trusted                                                                     & 1.0         & 0.0         & 1.0         & 0.0         & 1.0         & 0.0         \\  
                           & SPARAM                                                                         & 0.27        & 0.72        & 0.0         & 1.0         & 0.0         & 1.0         \\  
                           & PRNG                                                                    & 0.0         & 1.0         & 0.0         & 1.0         & 0.0         & 1.0         \\  
                           & HASH                                                                          & 0.0         & 1.0         & 0.0         & 1.0         & 0.0         & 1.0         \\ \hline
SPHINCS+                    & Trusted                                                                     & 0.82        & 0.18        & 0.84        & 0.16        & 1.0         & 0.0         \\  
                           & SPARAM                                                                         & 0.09        & 0.91        & 0.0         & 1.0         & 0.0         & 1.0         \\ 
                           & PRNG                                                                    & 0.0         & 1.0         & 0.0         & 1.0         & 0.0         & 1.0         \\  
                           & HASH                                                                          & 0.0         & 1.0         & 0.0         & 1.0         & 0.0         & 1.0         \\ \hline

Rainbow                    & Trusted                                                                     & 0.81        & 0.19        & 0.81        & 0.19        & 1.0         & 0.0         \\  
                           & SPARAM                                                                         & 0.0         & 1.0         & 0.0         & 1.0         & 0.0         & 1.0         \\ 
                           & PRNG                                                                    & 0.0         & 1.0         & 0.0         & 1.0         & 0.0         & 1.0         \\ 
                           & HASH                                                                          & 0.2         & 0.8         & 0.2         & 0.8         & 0.33        & 0.67        \\ \hline \hline
\end{tabular}
}

\label{tab:pcModel}
\end{center}
\end{table}

\section{Related Work}
\label{sec:relatedwork}

\textit{Side-channel attacks}: PQC implementations for resource constrained embedded processors are subject to implementation-based attacks. By tracing branch instructions on AVR micro-controllers one can recover the secret key in the BLISS PQC signature ~\cite{espitau2017side}. Power side-channel attack was performed on Dilithium using an intermediate value as the side-channel~\cite{ravi2018side}. A single trace is used as a side-channel on PQC lattice-based encryption schemes to perform key recovery ~\cite{primas2017single}. Power side-channels on FPGA implementations of McEliece PQC was shown in~\cite{chen2015horizontal}. Differential power side-channels were revealed in PQC XMSS and SPHINCS  ~\cite{kannwischer2018differential}. Resource-efficient fault attacks on pqm4 implementations are shown by~\cite{ravi2019exploiting}. These attacks show that PQC implementations on embedded platforms are vulnerable to side-channel attacks.

\textit{Algorithm subversion attacks}: Cryptography implementations are subject to code subversion attacks creating weaker signatures. Weaker RSA signature implementations are prone to  fault attacks~\cite{boneh2001importance}. Jafarholi \textit{et. al.,}~\cite{jafargholi2015tamper} present non-malleable codes as counter-measures for cryptographic algorithm subversion attacks. Cappos \textit{et. al.}~\cite{cappos2008look} show that software package managers are vulnerable to code subversion attacks when downloaded via untrusted channels. Subverting programs by accessing the disk memory of a user device is shown in~\cite{Evilmaid10}. The FREAK exploit~\cite{freak} shows that vulnerable SSL applications producing weaker signatures can be retrieved via man-in-the-middle attacks over untrusted networks. 
The POODLE exploit~\cite{poodle} downgrades the version of TLS similarly. 

\section{Conclusion}
\label{sec:conclusion}
PQC based digital signatures will be adopted in several domains replacing classic digital signatures. This work investigates various types of subverted PQC signature implementations.
Algorithm subversion attacks weaken the PQC signatures and make it vulnerable to attacks revealing secret information. Thus, securing PQC implementation on resource-constrained devices is a key requirement to maintain their integrity. We use Greybox fuzz testing to generate quality seed inputs to maximize state space coverage of a PQC implementation. These seed inputs aid in capturing unique HPC signatures and make our ML-based detection model robust against algorithm subversion attacks. The scheme makes it difficult for an adversary to hide malicious subversions from HPC signatures generated using Greybox fuzzed inputs.

\bibliographystyle{IEEEtran}
\bibliography{main.bib}

\begin{IEEEbiography}[{\includegraphics[width=1.2in,height=1.5in,clip,keepaspectratio]{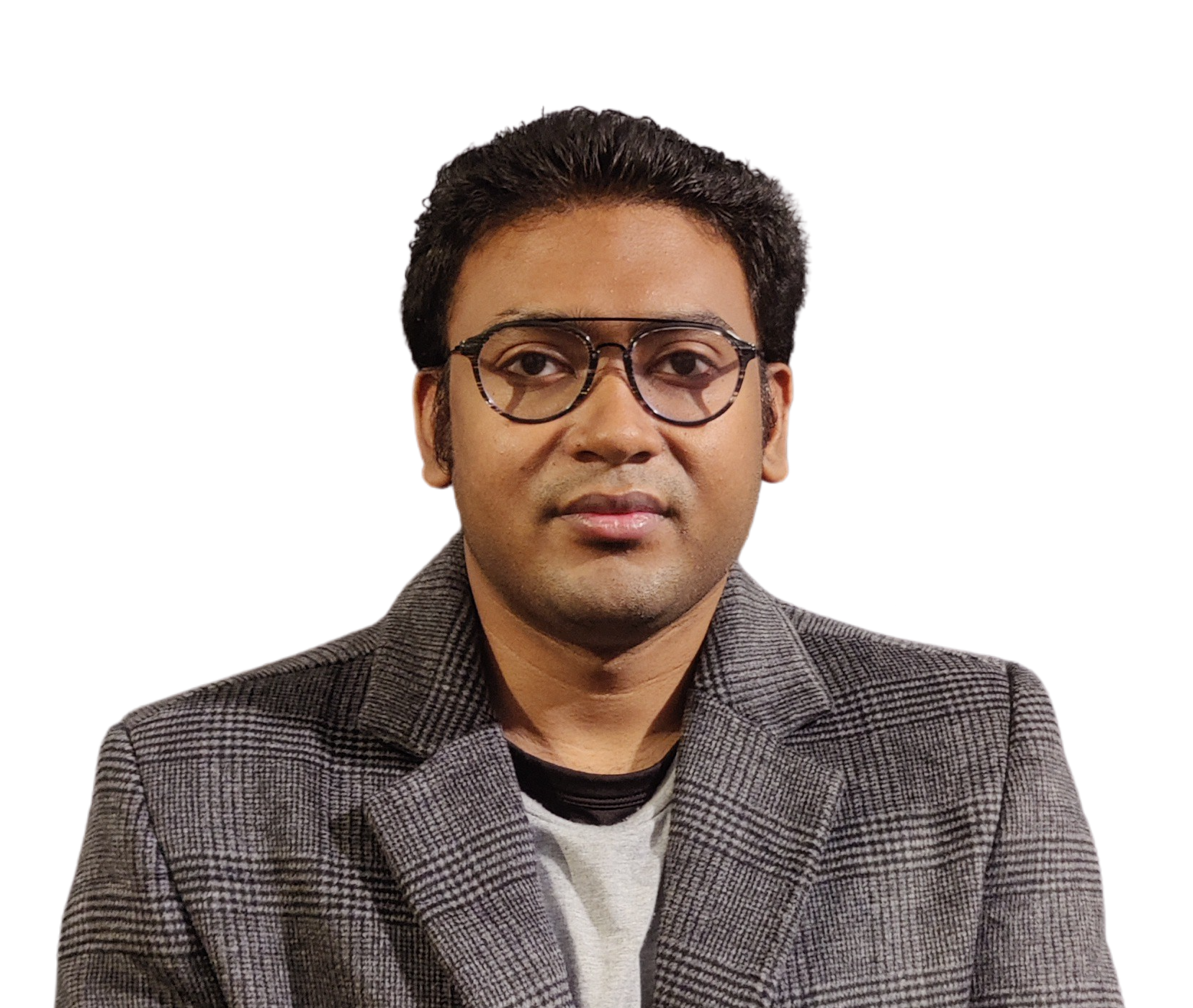}}]{Animesh Basak Chowdhury} is a Ph.D. candidate at the NYU Centre for Cybersecurity, where he works in the area of machine learning for Electronics Design Automation and security testing. He
received his MS in Computer Science from Indian Statistical Institute in 2016. Prior to joining the Ph.D. program, he spent three years as a researcher at Tata Research Development and Design Centre (TRDDC), India, where he was primarily working in the area of formal verification and security testing. He has won several awards and recognition in International Software Verification and Testing Competitions (SV-COMP, TEST COMP, and RERS-Challenge).

\end{IEEEbiography}

\begin{IEEEbiography}[{\includegraphics[width=1.2in,height=1.5in,clip,keepaspectratio]{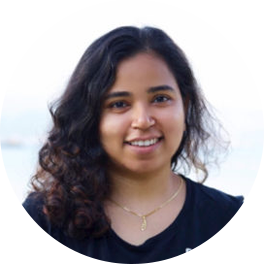}}]{Anushree Mahapatra} is a research engineer at Innatera Systems, Netherlands where she primarily works on electronics design automation problems of mapping Neural networks directly on hardware. She worked as post-doctoral researcher for a year at NYU Tandon School of Engineering in the areas like High-level synthesis security and Hardware Performance Counters based security. She received her Ph.D. from Hong Kong Poly-technique University in 2018 and MS from Nanyang Technological University in 2013. Her research interests include High level synthesis, design space exploration, machine learning and SoC security. Post completion of Ph.D., she spent a year in industry and worked on recommender systems.

\end{IEEEbiography}

\begin{IEEEbiography}[{\includegraphics[width=1.1in,height=1.5in,clip,keepaspectratio]{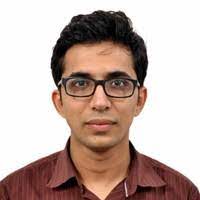}}]{Deepraj Soni} is a Ph.D. candidate at the NYU Tandon School of Engineering, where he works on hardware implementation, and evaluation and security of post-quantum cryptographic algorithms. He received his M.Tech from the Department of Electrical Engineering at the Indian Institute of Technology in Bombay (IIT-B). His thesis focused on developing a framework for a hardware-software co-simulator and neural network implementation on an FPGA. After graduation, Deepraj worked as a design engineer in the semiconductor division of Samsung and SanDisk. At Samsung, he was responsible for the design and architecture of image processing IPs, such as region segmentation and Embedded CODEC. He also had charge of communication IPs, such as FFT/IFFT, Time \& Frequency Deinterleaving and Demapper for canceling noise. At SanDisk, Deepraj helped in the development of System-On-Chip (SoC) level design for the memory controller.

\end{IEEEbiography}

\begin{IEEEbiography}[{\includegraphics[width=1.1in,height=1.25in,clip,keepaspectratio]{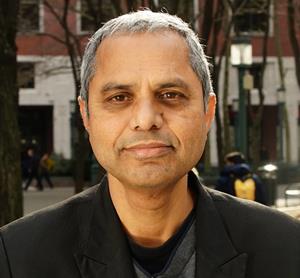}}]{Ramesh Karri} 
  is a Professor of ECE at New York University. He co-directs the NYU Center 
for Cyber Security (http://cyber.nyu.edu). He founded the Embedded Systems Challenge (https://csaw.engineering.nyu.edu/esc), the annual 
red team blue team event. He co-founded Trust-Hub (http://trust-hub.org). 
Ramesh Karri has a Ph.D. in Computer Science and 
Engineering, from the UC San Diego and a B.E in ECE from Andhra University. His 
research and education activities in hardware cybersecurity include trustworthy 
ICs; processors and cyber-physical systems; security-aware computer-aided 
design, test, verification, validation, and reliability; nano meets security; 
hardware security competitions, benchmarks, and metrics; biochip security; 
additive manufacturing security. He  published over 250 articles in leading 
journals and conference proceedings. Karri's work on hardware cybersecurity 
received best paper nominations (ICCD 2015 and DFTS 2015) and awards (ACM TODAES 
2018, ITC 2014, CCS 2013, DFTS 2013 and VLSI Design 2012). He received the 
Humboldt Fellowship and the NSF CAREER Award. He is the editor-in-chief of ACM JETC and serve(d)s  
on the editorial boards of IEEE and ACM Transactions (TIFS, TCAD, 
TODAES, ESL, D\&T, JETC). He was an IEEE Computer Society Distinguished 
Visitor (2013-2015). He served on the Executive Committee of the IEEE/ACM 
DAC leading the Security\@DAC initiative (2014-2017). 
He served as program/general chair of conferences and serves on program committees. 
He is a Fellow of the IEEE for leadership and contributions to Trustworthy Hardware.
\end{IEEEbiography}

\end{document}